\documentclass[groupedaddress, superscriptaddress, citesort]{revtex4}
\usepackage{amssymb, amsmath}
\usepackage{graphicx}

\def\be{\begin{equation}}
\def\ee{\end{equation}}
\def\ba{\begin{array}}
\def\ea{\end{array}}

\begin{document}
\baselineskip=18pt

\title {The Geometry of Quantum Coherence for Two Qubit $X$ States}
\author{Yao-Kun Wang}
\affiliation{College of Mathematics,  Tonghua Normal University, Tonghua, Jilin 134001, China}
\author{Lian-He Shao}
\affiliation{1College of Computer Science, Shaanxi Normal University, Xi¡¯an 710062, China}

\author{Yun-Ran Zhang}
\affiliation{Institute of Physics, Chinese Academy of Sciences, Beijing 100190, China}

\begin{abstract}
We plot the geometry of several distance-based quantifiers of coherence for Bell-diagonal states.
We find that along with both $l_{1}$ norm and relative entropy of coherence changes continuously from zero to one, their surfaces move from the separable regions to the entangled regions. Based on this fact,  it is more illuminating to use an intuitive geometry to explain quantum states with nonzero coherence can be used for entanglement creation, rather than the other way around.
We find the necessary and sufficient conditions that quantum discord of Bell-diagonal states equal to its relative entropy of coherence and depict the surfaces of the equality.
We give surfaces of relative entropy of coherence for $X$ states.
We show the surfaces of dynamics of relative entropy of coherence for Bell-diagonal states under local nondissipative channels and find that all coherence under local nondissipative channels decrease.
\end{abstract}

\maketitle

\section{Introduction}

Quantum coherence originates in the superposition of quantum states. Different from quantum entanglement and other quantum correlations, quantum coherence, regarded as a physical resource\cite{sashki,aberg,Baumgratz}, is an essential ingredient in quantum information processing\cite{Bagan,Jha,Kammerlander}, quantum metrology\cite{Giovannetti,Demkowicz,Giovannetti1}, quantum optics\cite{Glauber,Sudarshan,Mandel}, nanoscale thermodynamics\cite{Narasimhachar,Oppenheim,Lostaglio,Lostaglio1,Vazquez,Wacker}  and quantum biology\cite{Plenio,Rebentrost,Lloyd,Li,Huelga}.
Recently, a rigorous framework to quantify coherence has been proposed\cite{Baumgratz}. The framework is composed of  four conditions. Based on this framework, a number of quantum coherence measures, such as the $l_{1}$ norm of coherence\cite{Baumgratz}, the relative entropy of coherence\cite{Baumgratz}, trace norm of coherence\cite{shao},  Tsallis relative $\alpha$ entropies\cite{Rastegin} and Relative R\'{e}nyi $\alpha$ monotones\cite{Chitambar}, have been put forward.
With the coherence measures, many properties of quantum coherence, such as the relations between quantum coherence and quantum correlations\cite{Ma,Radhakrishnan,Streltsov,Yao,Xi}, the relative entropy of coherence decreasing strictly for all nontrivial evolutions preserving in the dynamics of coherence\cite{Bromley}, the freezing phenomenon of coherence\cite{Bromley,Yu}, were investigated.

The geometry of Bell-diagonal states, including the subsets of separable and classical subsets, can be depicted in three dimensions\cite{Horodecki,Horodecki1}. Level surfaces of entanglement and nonclassical measures can be plotted directly on this three-dimensional geometry. For this simple case, the result show a complete picture of the structure of entanglement and nonclassicality. It is more illuminating to use this picture to explain how measures of entanglement and nonclassicality change rather than the other way around\cite{Lang}. What is more, a large number of related research appear, such as the level surfaces of quantum discord for a class of two-qubit states\cite{Li1}, geometry of one-way Information deficit of a class of two-qubit states\cite{wang}, the surfaces of constant quantum discord and super-quantum discord for Bell-diagonal state\cite{wang1}, geometric illustration of Bell-diagonal states steerable by two projective measurements\cite{quan}.

In this article, we calculate several distance-based quantifiers of coherence for Bell-diagonal states and plot their geometry. We plot the geometry both $l_{1}$-norm of coherence and relative entropy of coherence for Bell-diagonal states in separable and entangled regions. We give the surfaces that quantum discord equal to relative entropy of coherence for Bell-diagonal states. We depict the geometry of relative entropy of coherence for $X$ states. We show the geometry of dynamics of relative entropy of coherence for Bell-diagonal states under local nondissipative channels.

\section{Geometry of several distance-based quantifiers of coherence for Bell-diagonal states}

First, we will introduce the form two-qubit of Bell-diagonal states. We can write Bell-diagonal states as
\begin{eqnarray}
\rho^{ab}=\frac{1}{4}(I\otimes I+\sum_{i=1}^3c_i\sigma_i\otimes\sigma_i), \label{state1}
\end{eqnarray}
where  $\{\sigma_i\}_{i=1}^3$ are the standard Pauli matrices. In the computational basis ${|00\rangle, |01\rangle, |10\rangle, |11\rangle}$, its density matrix is
\begin{eqnarray}\label{state9}
\rho = \frac{1}{4} \left(
\begin{array}{cccc}
1+c_3
& 0 & 0 & c_1 -c_2 \\
0 & 1-c_3 & c_1+c_2 & 0 \\
0 & c_1 +c_2 & 1-c_3
& 0 \\
c_1 -c_2 & 0 & 0 & 1+c_3
\end{array}
\right) \,.
\end{eqnarray}
where $c_{1}, c_{2}, c_{3}\in[-1,1]$.

We will give the geometry of several distance-based quantifiers of coherence for Bell-diagonal states, such as $l_{1}$-norm of coherence $C_{l_{1}}$\cite{Baumgratz}, trace distance of $C_{tr}$\cite{shao}, relative entropy of coherence $C_{r}$\cite{Baumgratz}.  As Bell-diagonal states is $X$ states, $C_{l_{1}}=C_{tr}$\cite{Rana}. We will give the formula of coherence above for Bell-diagonal states as follow:

\begin{eqnarray}\label{l1}
C_{l_{1}}(\rho)=\sum_{i\neq j}|\rho_{i,j}|=\frac{1}{2}(|c_{1}-c_{2}|+|c_{1}+c_{2}|).
\end{eqnarray}

\begin{eqnarray}\label{relative}
C_{r}(\rho)&=&S(\rho_{diag})-S(\rho)\nonumber\\
&=&\frac{1}{4}(1-c_{1}-c_{2}-c_{3})\log\frac{1}{4}(1-c_{1}-c_{2}-c_{3})\nonumber\\
 &+&\frac{1}{4}(1-c_{1}+c_{2}-c_{3})\log\frac{1}{4}(1-c_{1}+c_{2}-c_{3})\nonumber\\
 &+&\frac{1}{4}(1+c_{1}-c_{2}+c_{3})\log\frac{1}{4}(1+c_{1}-c_{2}+c_{3})\nonumber\\
 &+&\frac{1}{4}(1+c_{1}+c_{2}+c_{3})\log\frac{1}{4}(1+c_{1}+c_{2}+c_{3})\nonumber\\
 &+&2-\frac{1+c_{3}}{2}\log(1+c_{3})-\frac{1-c_{3}}{2}\log(1-c_{3}).
\end{eqnarray}



Next, we will give the geometry of the measures of coherence above. The geometry between $l_{1}$-norm of coherence and relative entropy of coherence are basically the same except the shape of the parallelogram[see Fig 1. and Fig 2.]. The orange tetrahedron is the set of valid Bell-diagonal states. The green octahedron is the set of separable Bell-diagonal states. There are four entangled regions outside octahedron\cite{Lang}. The level surface like the tubes running along the Cartesian axes $C_{3}$. The tubes are cut off by the state tetrahedron at their ends. As coherence decreases, the tubes collapse to the Cartesian axes. As coherence increases, the tube structure is obscured. From  Eq. (\ref{l1}) and Eq. (\ref{relative}), one can know that coherence equal to $0$ if and only if $c_{1}=c_{2}=0$, and the geometry of coherence is the Cartesian axes $C_{3}$ that is in the set of separable Bell-diagonal states(the green octahedron) completely. When coherence is tiny, surfaces start to move to entangled region[see (a) of the Fig 1. and Fig. 2]. As coherence increase, the surfaces distribute in regions both separable and entangled[see (b) of the Fig. 1 and Fig. 2]. When coherence approaches 1, its surfaces almost distribute in entangled regions[see (c) of the Fig 1. and Fig. 2]. When coherence equal to $1$, its surfaces is the four vertexes of the tetrahedron which is four Bell states.

We also plot the geometry both Tsallis $\alpha$ divergence \cite{Rastegin} and  Relative R\'{e}nyi $\alpha$ monotones \cite{Chitambar} for $\alpha=1/2, 3/2, 2$. We find their geometry is extremely similar to the geometry of relative entropy of coherence. Here we won't repeat.


\begin{figure}[h]
\begin{center}
\scalebox{1.0}{(a)}{\includegraphics[width=4.5cm]{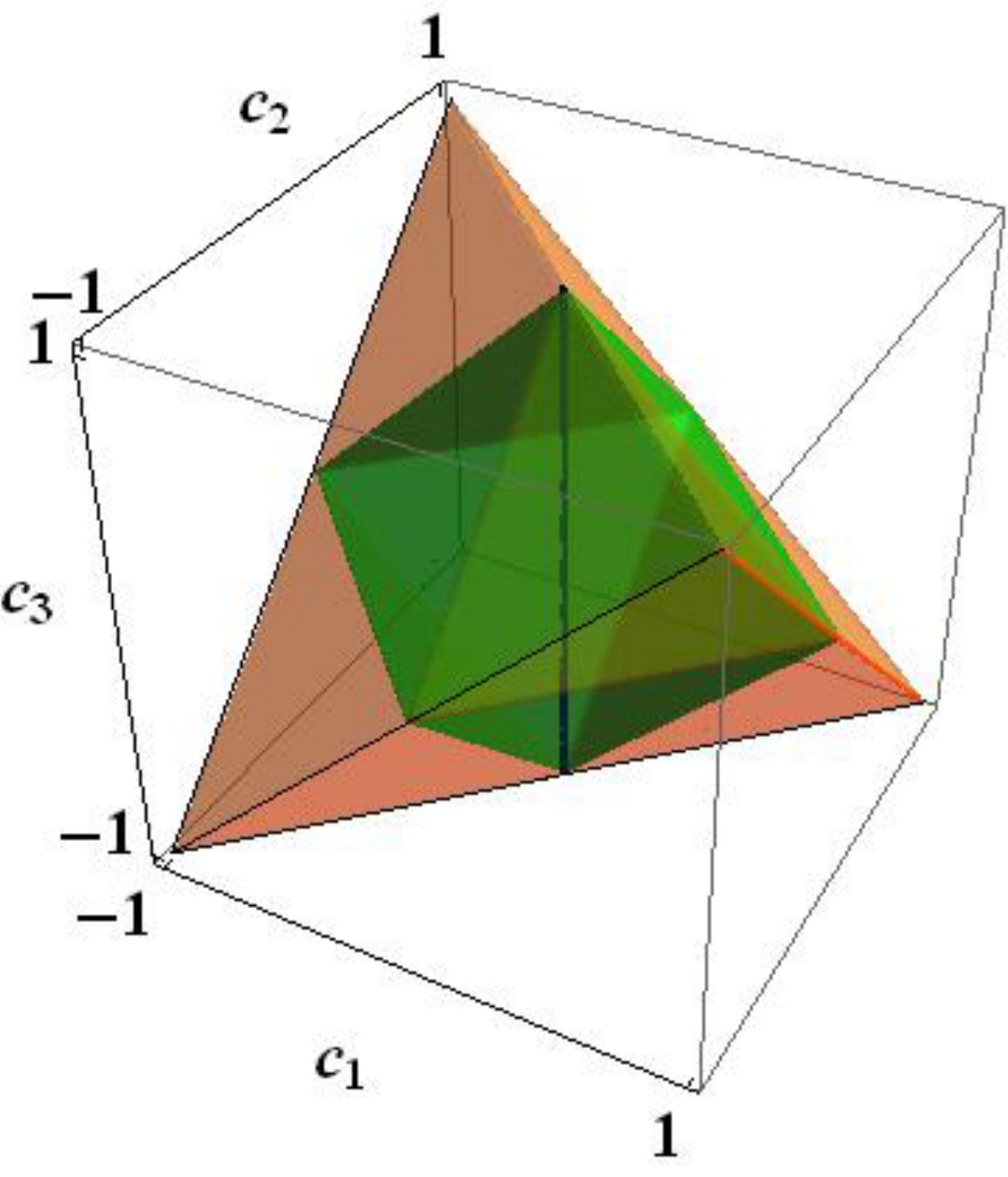}}
\scalebox{1.0}{(b)}{\includegraphics[width=4.5cm]{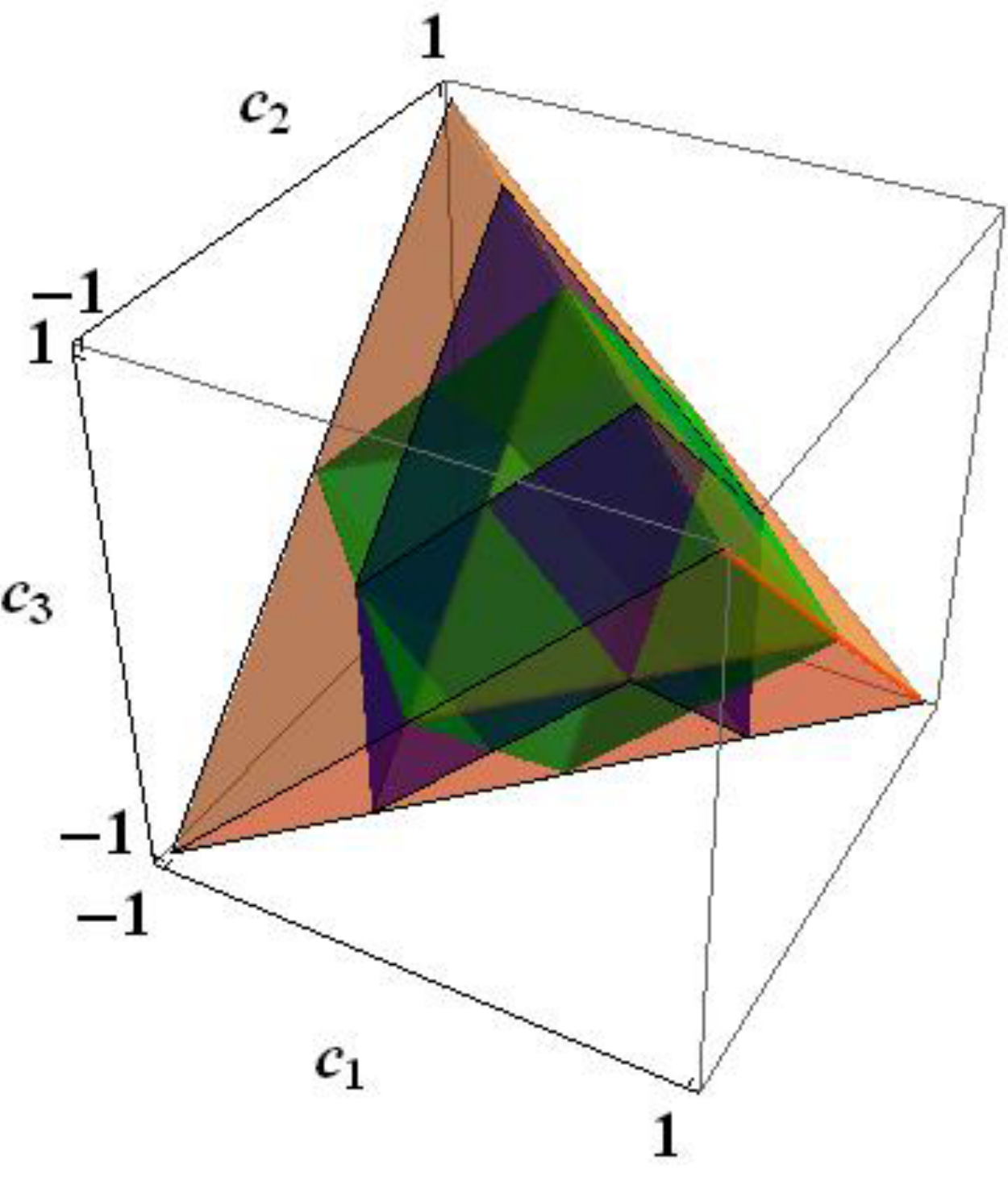}}
\scalebox{1.0}{(c)}{\includegraphics[width=4.5cm]{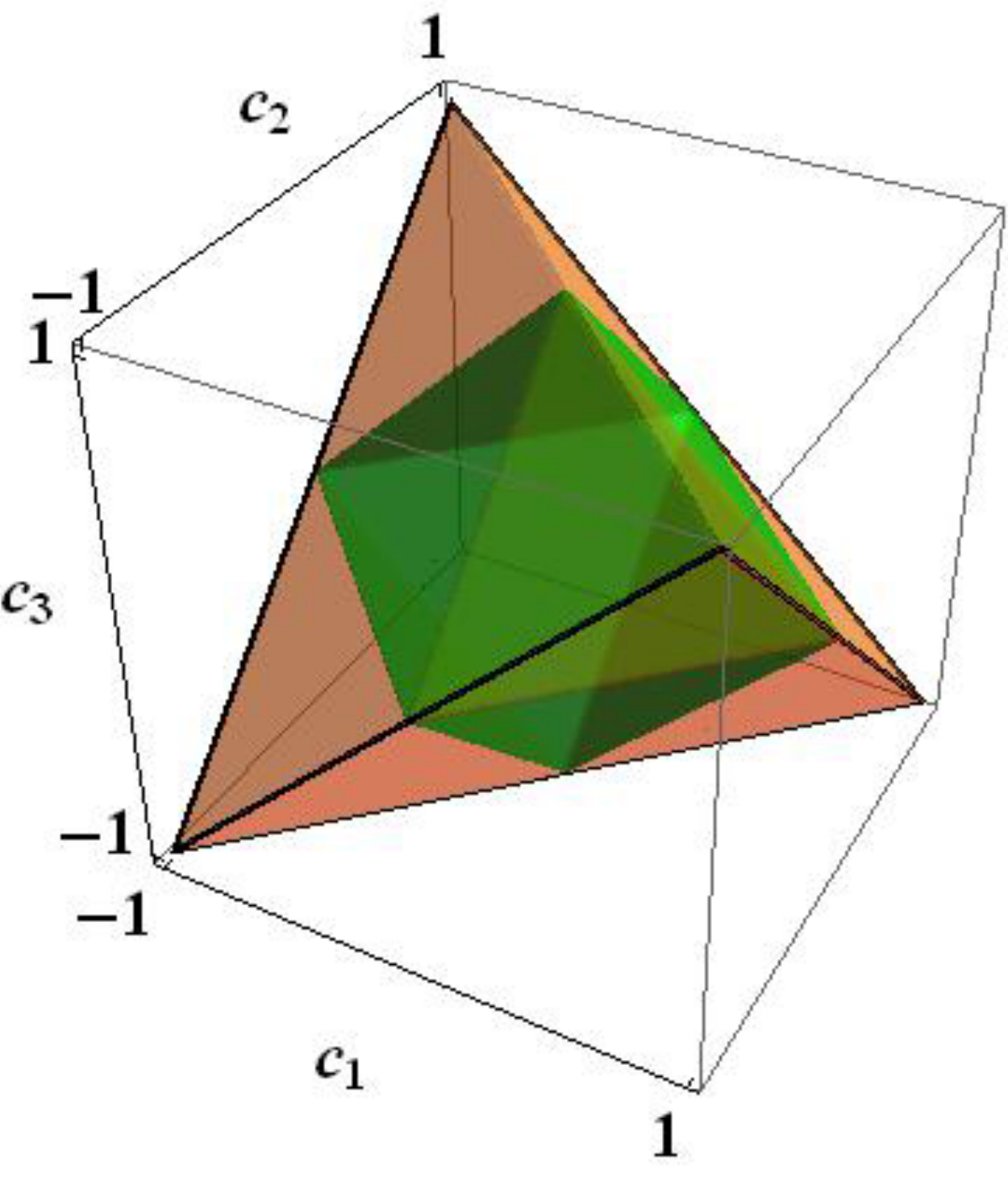}}
\caption{Surfaces of constant $l_{1}$-norm of coherence $C_{l_{1}}$ for Bell-diagonal states (Blue surfaces): (a) $C_{l_{1}}=0.001$; (b) $C_{l_{1}}=0.5$; (c) $C_{l_{1}}=0.99$.}
\end{center}
\end{figure}

\begin{figure}[h]
\begin{center}
\scalebox{1.0}{(a)}{\includegraphics[width=4.5cm]{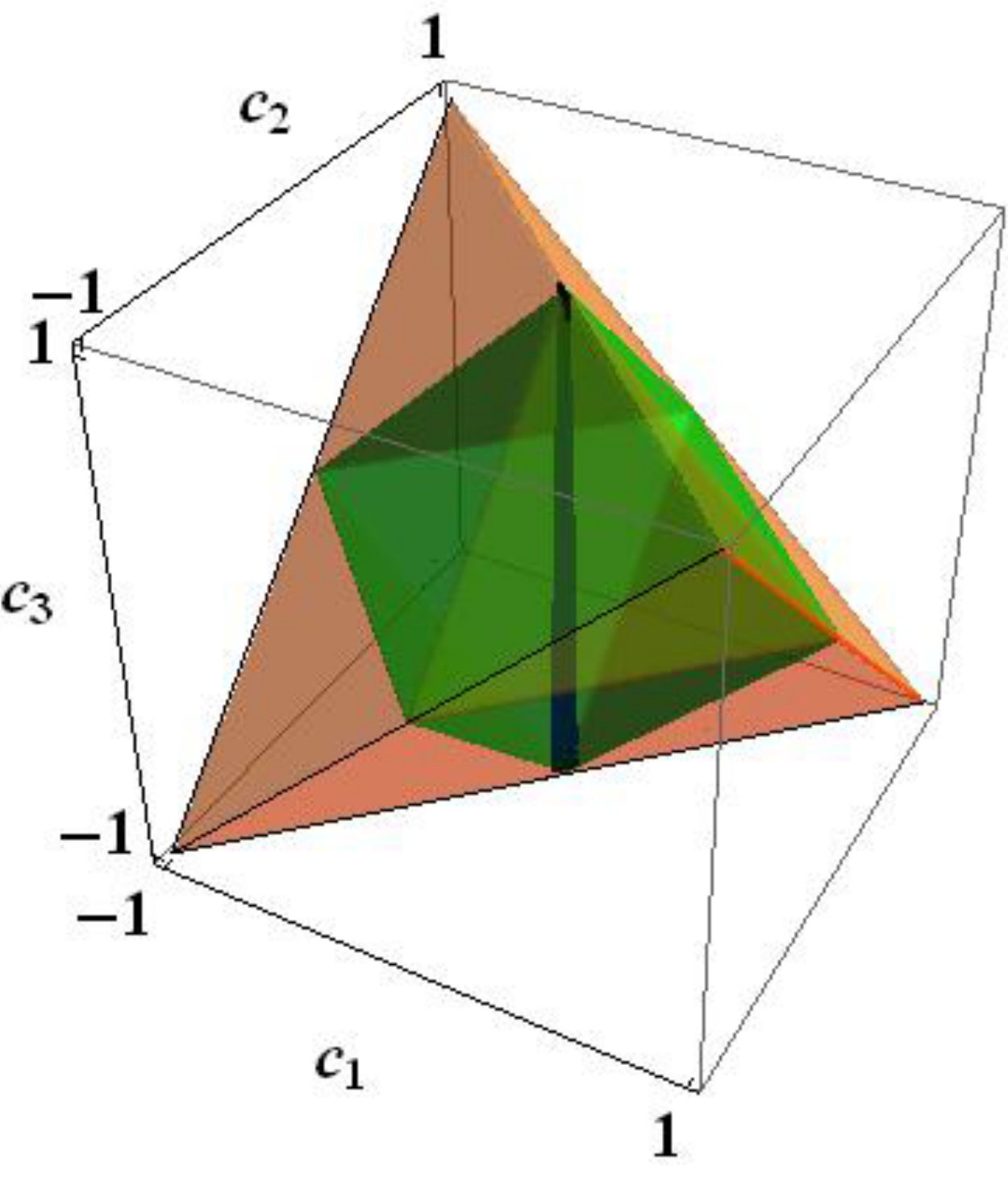}}
\scalebox{1.0}{(b)}{\includegraphics[width=4.5cm]{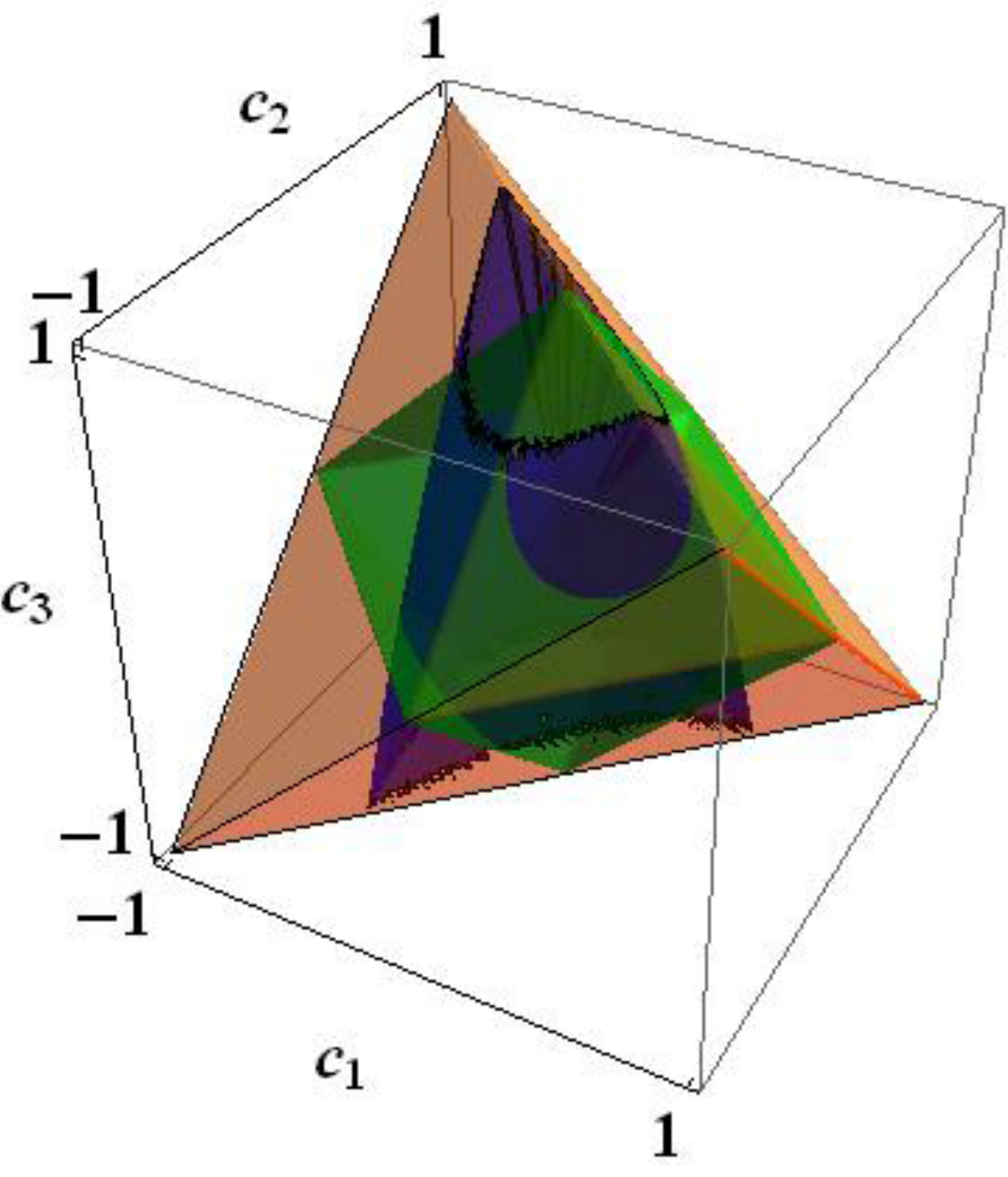}}
\scalebox{1.0}{(c)}{\includegraphics[width=4.5cm]{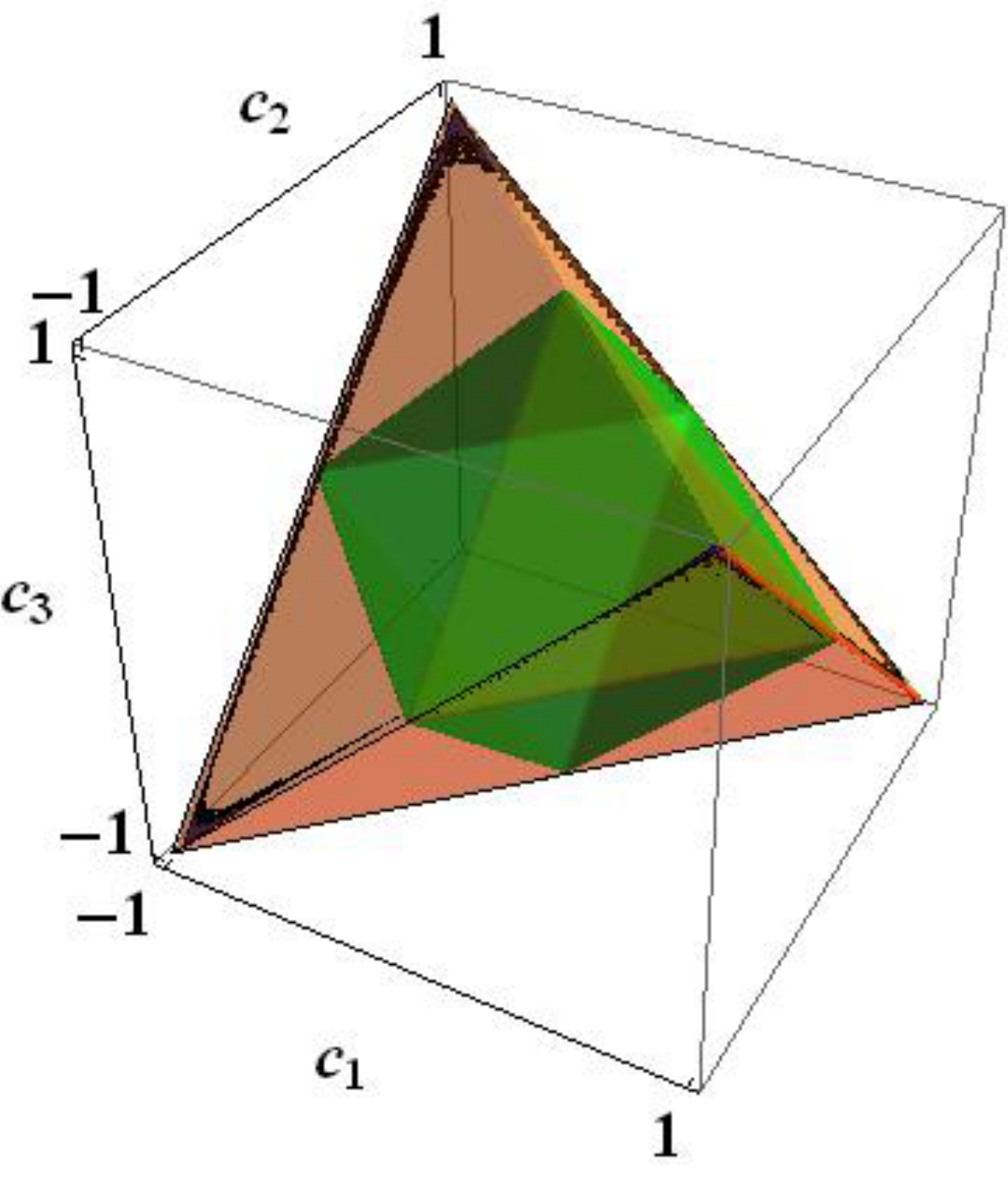}}
\caption{Surfaces of constant relative entropy of coherence for Bell-diagonal states (Blue surfaces): (a) $C_{r}(\rho)=0.001$; (b) $C_{r}(\rho)=0.2$; (c) $C_{r}(\rho)=0.9$.}
\end{center}
\end{figure}

We will give the surfaces that quantum discord equal to relative entropy of coherence for Bell-diagonal states. Quantum discord of Bell-diagonal states is given by\cite{luo}
\begin{eqnarray}
D(\rho)&=&\frac{1}{4}(1-c_{1}-c_{2}-c_{3})\log\frac{1}{4}(1-c_{1}-c_{2}-c_{3})\nonumber\\
 & &+\frac{1}{4}(1-c_{1}+c_{2}-c_{3})\log\frac{1}{4}(1-c_{1}+c_{2}-c_{3})\nonumber\\
 & &+\frac{1}{4}(1+c_{1}-c_{2}+c_{3})\log\frac{1}{4}(1+c_{1}-c_{2}+c_{3})\nonumber\\
 & &+\frac{1}{4}(1+c_{1}+c_{2}+c_{3})\log\frac{1}{4}(1+c_{1}+c_{2}+c_{3})\nonumber\\
 & &+2-\frac{1+c}{2}\log(1+c)-\frac{1-c}{2}\log(1-c),\nonumber
\end{eqnarray}
where $c=max\{|c_{1}|,|c_{2}|,|c_{3}|\}$.
By comparison with Eq. (\ref{relative}), quantum discord equal to relative entropy of coherence for Bell-diagonal states if and only if $c_{3}=max\{|c_{1}|,|c_{2}|,|c_{3}|\}$. In Fig. 3 two blue tubes above and below the orange cross tube represent the surfaces of $C_{r}(\rho)=D(\rho)=0.05$.

\begin{figure}[ht]
\includegraphics[scale=0.4]{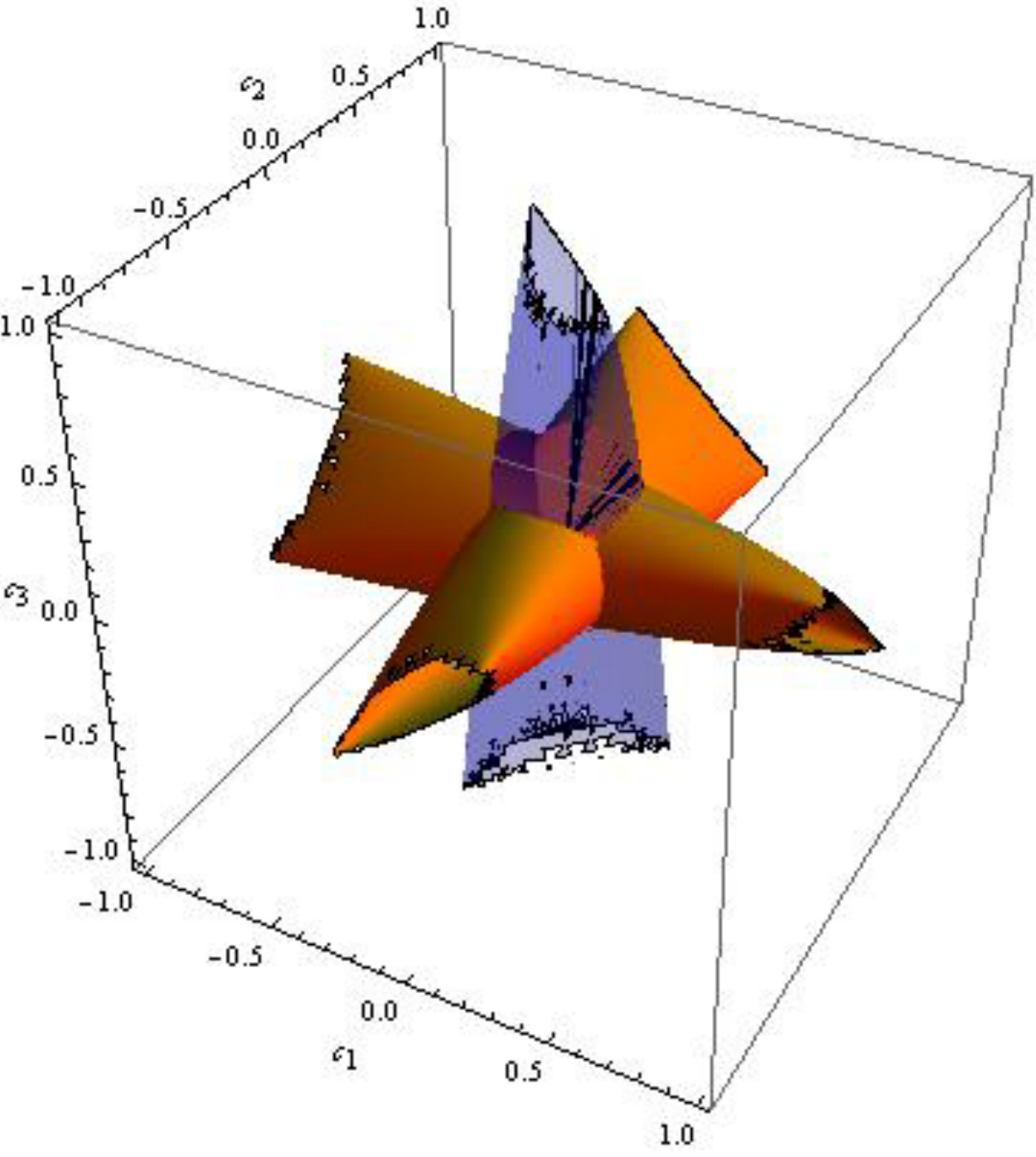}
\caption{(Color online) Surfaces of $C_{r}(\rho)=D(\rho)=0.05$ for Bell-diagonal states (Two Blue tubes).}
\end{figure}

\section{Geometry of relative entropy of coherence for $X$ states}

We will introduce the form of two qubit $X$ states. By using proper local unitary transformations,  we can write $\rho^{ab}$ as
\begin{eqnarray}
\rho^{ab}=\frac{1}{4}(I\otimes I+\textbf{r}\cdot\sigma\otimes I+I\otimes\textbf{s}\cdot\sigma+\sum_{i=1}^3c_i\sigma_i\otimes\sigma_i), \label{state1}
\end{eqnarray}
where \textbf{r} and \textbf{s} are Bloch vectors and $\{\sigma_i\}_{i=1}^3$ are
the standard Pauli matrices. When \textbf{r}=\textbf{s}=\textbf{0},
$\rho$ reduces to the two-qubit Bell-diagonal states.
Then, we assume that the Bloch vectors are in $z$ direction,  that is,  $\textbf{r}=(0, 0, r)$,  $\textbf{s}=(0, 0, s)$. The state in Eq. (\ref{state1}) turns into the following form
\begin{eqnarray}
\rho^{ab}=\frac{1}{4}(I\otimes I+r\sigma_{3}\otimes I+I\otimes s\sigma_{3}+\sum_{i=1}^3c_i\sigma_i\otimes\sigma_i).\nonumber
\end{eqnarray}
In the computational basis ${|00\rangle, |01\rangle, |10\rangle, |11\rangle}$, its density matrix is
\begin{eqnarray}\label{state3}
\rho = \frac{1}{4} \left(
\begin{array}{cccc}
1+r+s+c_3
& 0 & 0 & c_1 -c_2 \\
0 & 1+r-s-c_3 & c_1+c_2 & 0 \\
0 & c_1 +c_2 & 1-r+s-c_3
& 0 \\
c_1 -c_2 & 0 & 0 & 1-r-s+c_3
\end{array}
\right) \,.
\end{eqnarray}
From Eq. (4) in \cite{chen}, after some algebraic calculations, we can obtain that parameters $x, y, s, u, t$ in \cite{chen}
can be substituted for $r, s, c_{1}, c_{2}, c_{3}$ of the $X$ states in Eq. (\ref{state3}) successively and $r, s, c_{1}, c_{2}, c_{3}\in[-1,1]$.

Relative entropy of coherence for $X$ states is given by
\begin{eqnarray}
C_{r}(\rho)&=&S(\rho_{diag})-S(\rho)\nonumber\\
       &=&\frac{1}{4}(1-c_3+\sqrt{(r-s)^2+(c_1+c_2)^2})\log(1-c_3+\sqrt{(r-s)^2+(c_1+c_2)^2})\nonumber\\
       & &+\frac{1}{4}(1-c_3-\sqrt{(r-s)^2+(c_1+c_2)^2})\log(1-c_3-\sqrt{(r-s)^2+(c_1+c_2)^2})\nonumber\\
       & &+\frac{1}{4}(1+c_3+\sqrt{(r+s)^2+(c_1-c_2)^2})\log(1+c_3+\sqrt{(r+s)^2+(c_1+c_2)^2})\nonumber\\
       & &+\frac{1}{4}(1+c_3-\sqrt{(r+s)^2+(c_1-c_2)^2})\log(1+c_3-\sqrt{(r+s)^2+(c_1-c_2)^2})\nonumber\\
       & &-\frac{1}{4}(1+r+s+c_3)\log(1+r+s+c_3)\nonumber\\
       & &-\frac{1}{4}(1+r-s-c_3)\log(1+r-s-c_3)\nonumber\\
       & &-\frac{1}{4}(1-r+s-c_3)\log(1-r+s-c_3)\nonumber\\
       & &-\frac{1}{4}(1-r-s+c_3)\log(1-r-s+c_3).
\end{eqnarray}

In Fig. 4 we plot surface of relative entropy of coherence for $X$ states. From Fig. 4, one can see that the level surface of coherence is similar to the case $r = s =0$ which is Bell-diagonal states above [see Fig. 4(a)]. The surface shrinks with the effect of $r$ and $s$ and the shrinking rate becomes larger with the increasing $|r|$ and $|s|$ [see Fig. 4(a), (b)]. For larger $r$ and $s$, the picture is moved up the plane $c_{3}=0$ [see Fig. 4(b)]. For larger coherence and small $r$ and $s$ [see Fig. 4(c)], the surface become fat and short. But for larger $r$ and $s$ [see Fig. 4(d)], the figure is moved up again and changes dramatically also.

\begin{figure}[h]
\raisebox{6.2em}{(a)}\includegraphics[height=3.90cm,width=3.90cm]{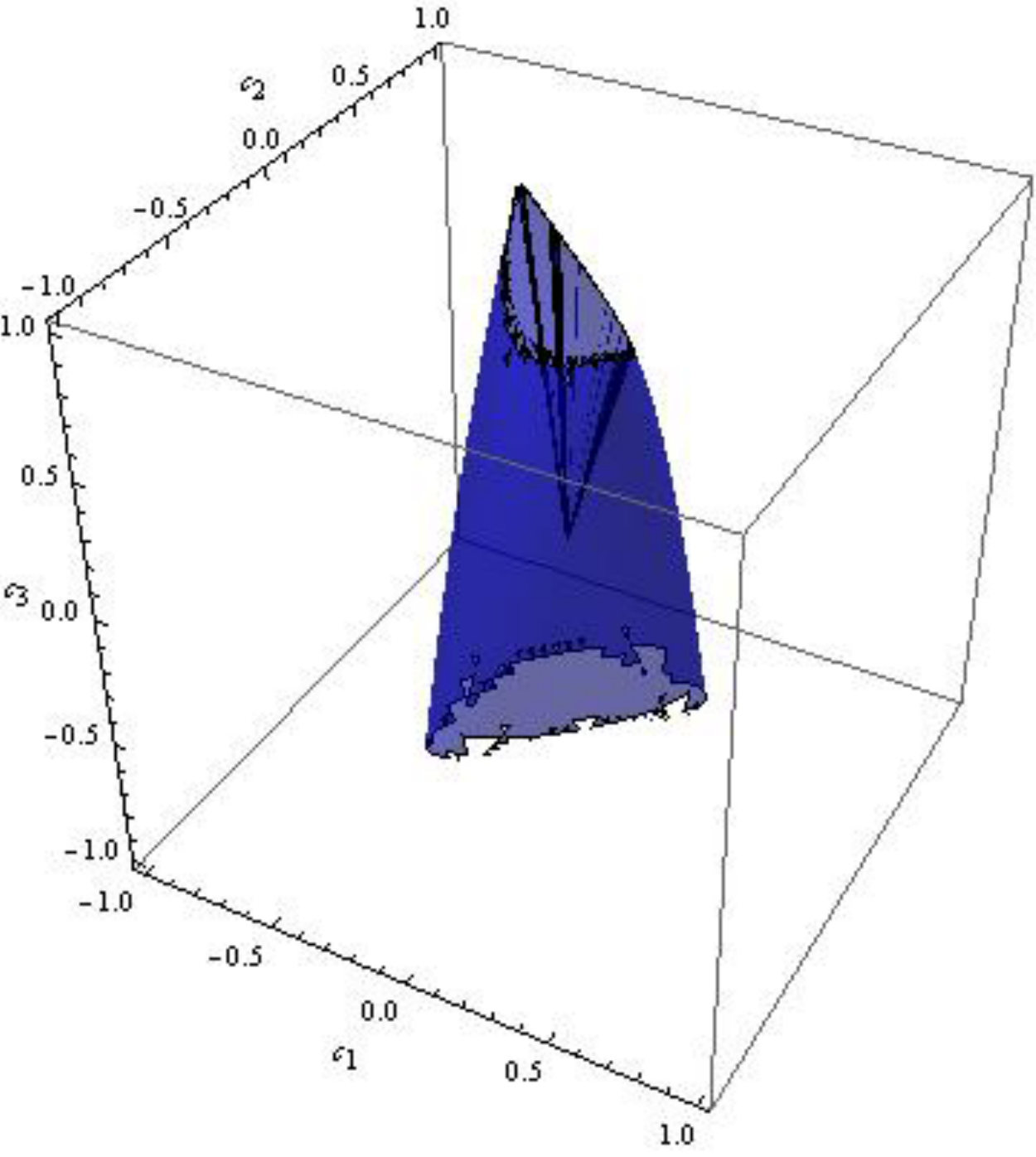}
\qquad
\raisebox{6.2em}{(b)}\includegraphics[height=3.90cm,width=3.90cm]{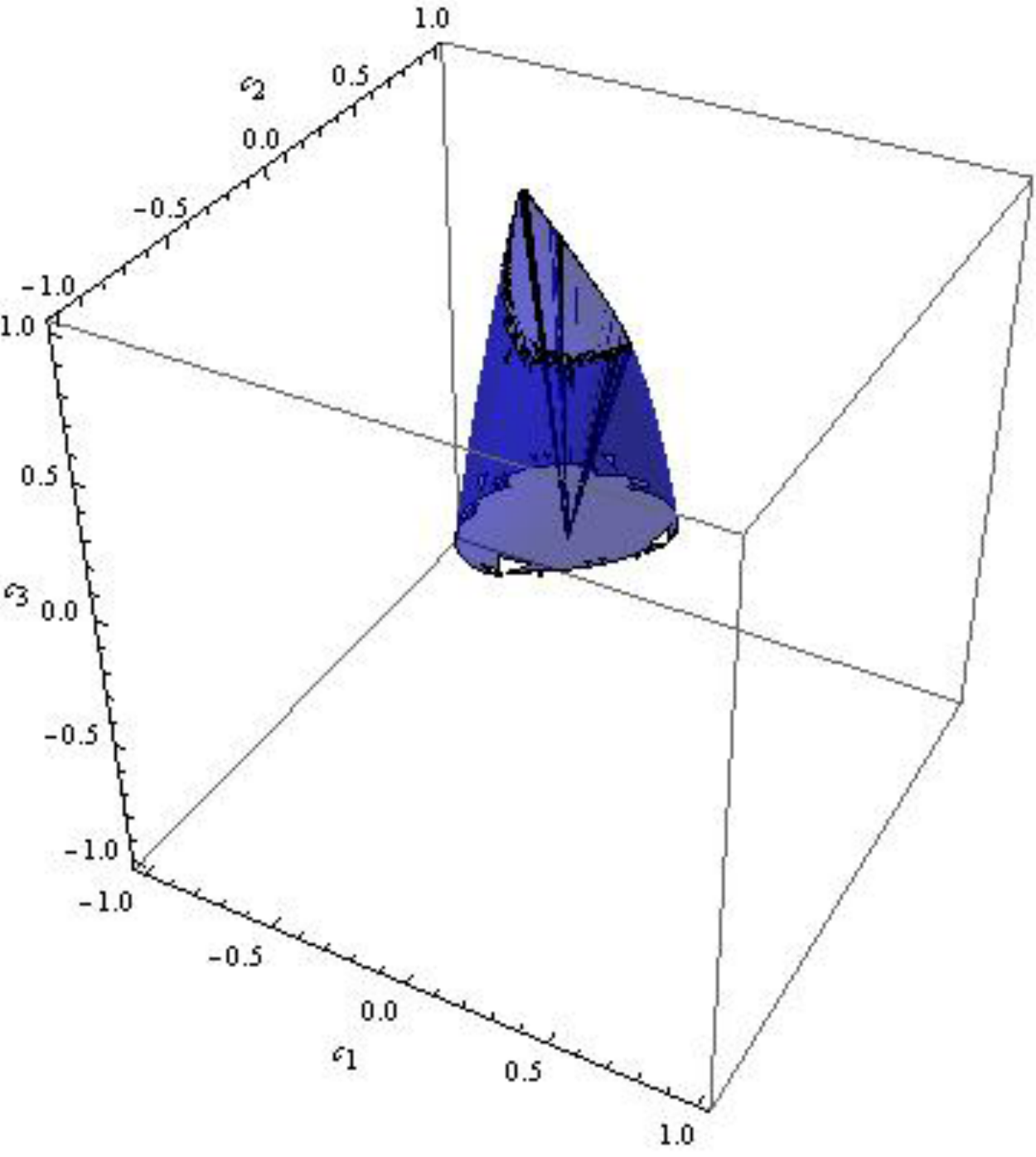}
\end{figure}
\begin{figure}[h]
\begin{center}
\raisebox{6.2em}{(c)}\includegraphics[height=3.90cm,width=3.90cm]{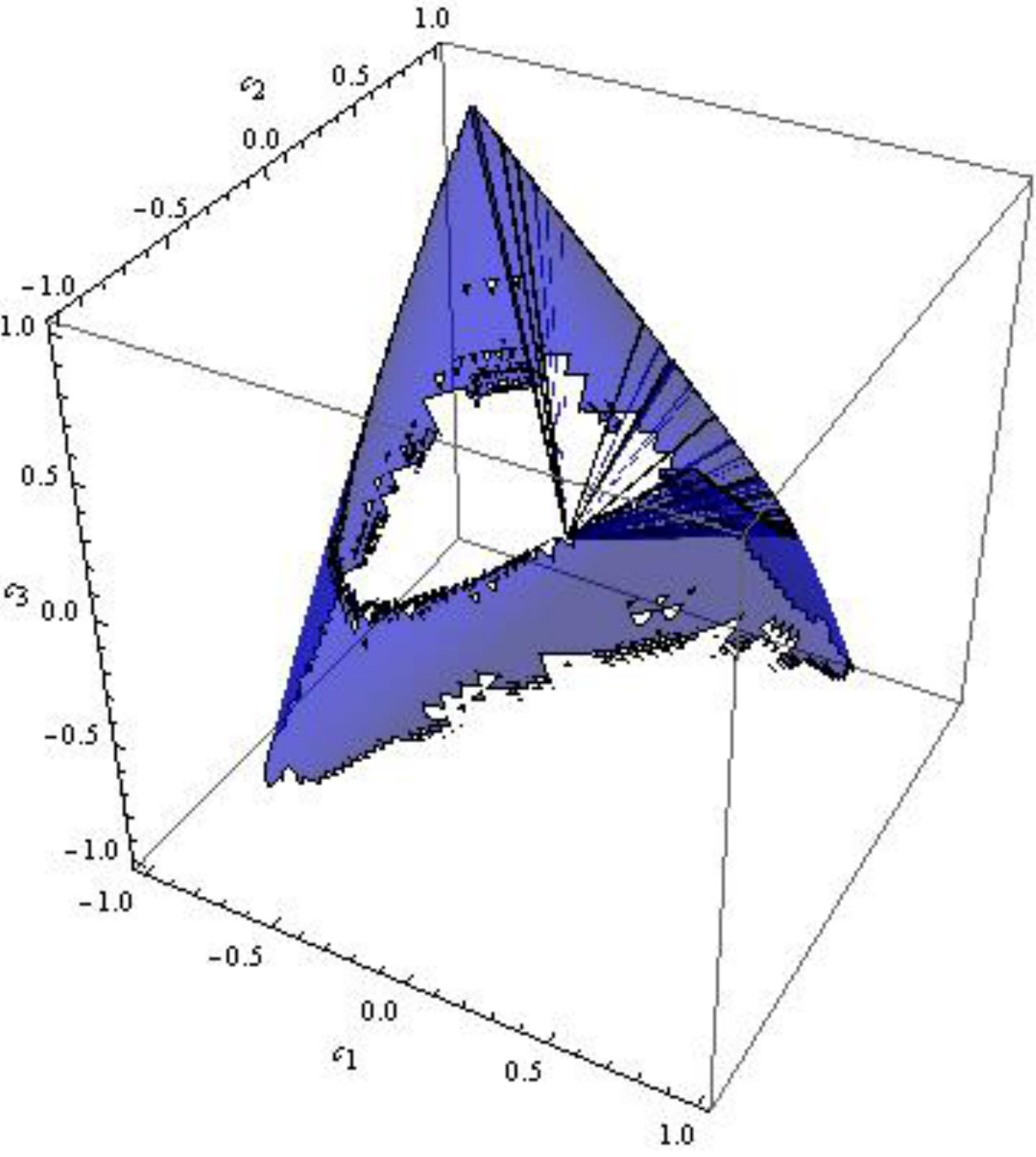}
\qquad
\raisebox{6.2em}{(d)}\includegraphics[height=3.90cm,width=3.90cm]{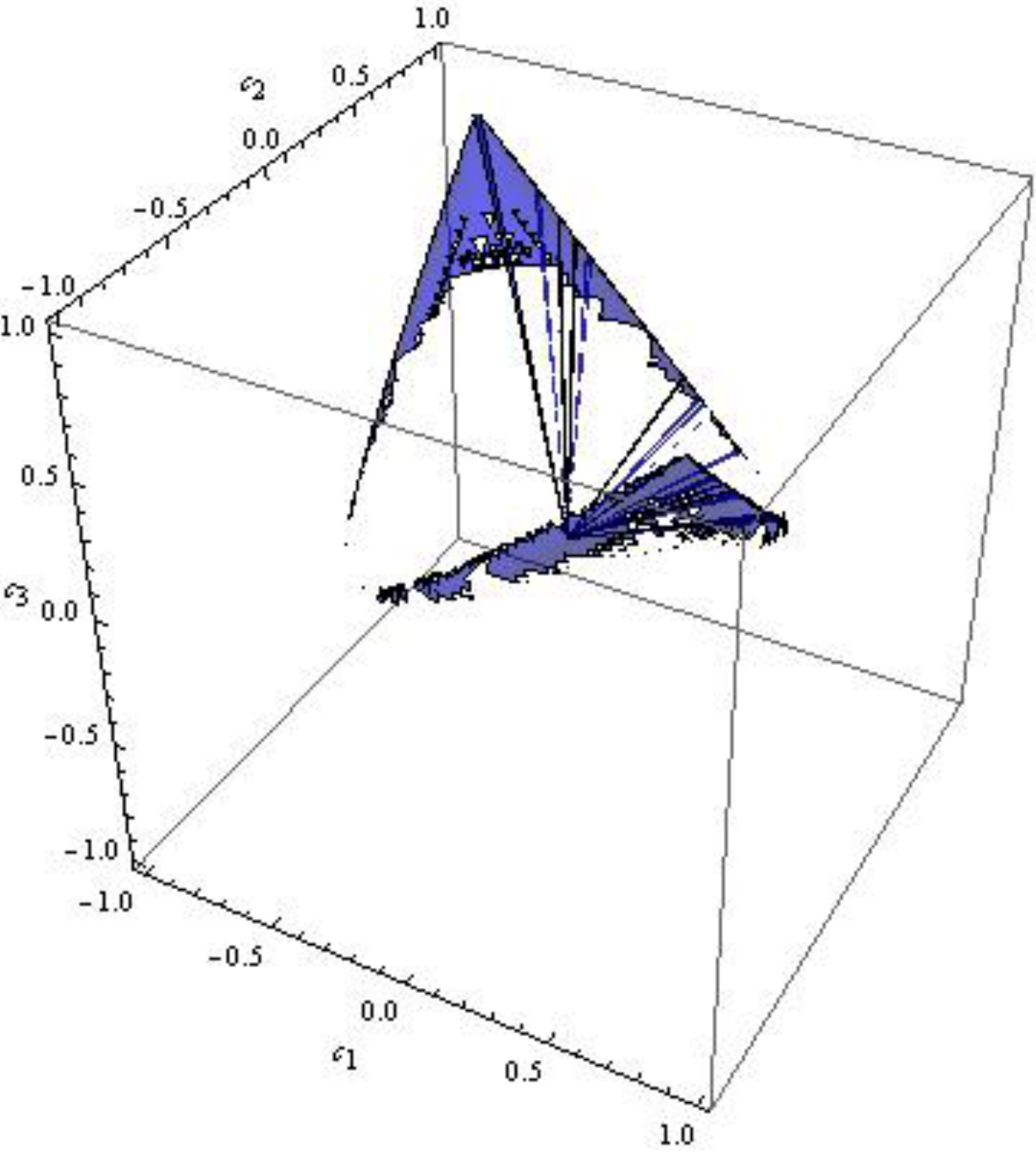}
\end{center}
\caption{(Color online) Surfaces of the constant of relative entropy of coherence for $X$ states(blue surface): (a)$r=s=0.1, C_{r}=0.1$; (b)$r=s=0.5, C_{r}=0.1$; (c)$r=s=0.1, C_{r}=0.5$; (d)$r=s=0.5, C_{r}=0.5$.}
\end{figure}

\section{Geometry of dynamics of relative entropy of coherence for Bell-diagonal states under local nondissipative channels.}
We will consider the system-environment interaction\cite{nielsen} through the evolution of a quantum state
$\rho$ under a trace-preserving quantum operation $\varepsilon(\rho)$,
\begin{equation}
\varepsilon(\rho) = \sum_{i,j} \left(E_i\otimes E_j\right) \rho \left(E_i \otimes E_j\right)^\dagger,\nonumber
\end{equation}
where $\{E_k\}$ is the set of Kraus operators associated to a decohering process of a single qubit,
with $\sum_k E_k^\dagger E_k = I$. We will use the Kraus operators in Table~\ref{t1} \cite{mont} to describe a variety of channels considered in this work.

\begin{table}
\begin{center}
\begin{tabular}{c c}
\hline \hline
 & $\textrm{Kraus operators}$                                         \\  \hline & \\
BF   & $E_0 = \sqrt{1-p/2}\, I , E_1 = \sqrt{p/2} \,\sigma_1$                        \\
 & \\
PF   & $E_0 = \sqrt{1-p/2}\, I , E_1 = \sqrt{p/2}\, \sigma_3$                        \\
 & \\
BPF & $E_0 = \sqrt{1-p/2}\, I , E_1 = \sqrt{p/2} \,\sigma_2$                        \\
 & \\
GAD   &
$E_0=\sqrt{p}\left(
\begin{array}{cc}
1 & 0 \\
0 & \sqrt{1-\gamma} \\
\end{array} \right) ,
E_2=\sqrt{1-p}\left(
\begin{array}{cc}
\sqrt{1-\gamma} & 0 \\
0 & 1 \\
\end{array} \right)$  \\
& \\
 & $E_1=\sqrt{p}\left(
\begin{array}{cc}
0 & \sqrt{\gamma} \\
0 & 0 \\
\end{array} \right) ,
E_3=\sqrt{1-p}\left(
\begin{array}{cc}
0 & 0 \\
\sqrt{\gamma} & 0 \\
\end{array} \right)$  \\ \hline \hline
\end{tabular}
\caption[table1]{Kraus operators for the quantum channels: bit flip (BF), phase flip (PF),
bit-phase flip (BPF), and generalized amplitude damping (GAD), where
$p$ and $\gamma$ are decoherence probabilities, $0<p<1$, $0<\gamma<1$.}
\label{t1}
\end{center}
\end{table}

The decoherence processes BF, PF, and BPF in Table~\ref{t1} preserve the Bell-diagonal form of the
density operator $\rho_{AB}$.
For the case of GAD, the Bell-diagonal form is kept for arbitrary $\gamma$
and $p=1/2$. In this situation, we can write the quantum operation $\varepsilon(\rho)$ as
\begin{equation}
\varepsilon(\rho_{AB})=\frac{1}{4}(I\otimes I+\sum_{i=1}^3c^\prime_i\sigma_i\otimes\sigma_i)\label{newstate},\nonumber
\end{equation}
where the values of the $c^\prime_1$, $c^\prime_2$, $c^\prime_3$
are given in Table~\ref{t2} \cite{mont}.
\begin{table}
\begin{center}
\begin{tabular}{c c c c}
\hline \hline
$\textrm{Channel}$ & $c^\prime_1$      & $c^\prime_2$     & $c^\prime_3$      \\ \hline
& & & \\
BF                 &  $c_1$            & $c_2 (1-p)^2$    & $c_3 (1-p)^2$     \\
& & & \\
PF                 &  $c_1 (1-p)^2$    & $c_2 (1-p)^2$    & $c_3$             \\
& & & \\
BPF                &  $c_1 (1-p)^2$    & $c_2$            & $c_3 (1-p)^2$     \\
& & & \\
GAD                &  $c_1 (1-p)$ & $c_2 (1-p)$ & $c_3 (1-p)^2$ \\ \hline \hline
\end{tabular}
\caption[table2]{Correlation functions for the quantum operations: bit flip (BF), phase flip (PF),
bit-phase flip (BPF), and generalized amplitude damping (GAD). For GAD, we
fixed $p=1/2$ and replaced $\gamma$ with $p$.}
\label{t2}
\end{center}
\end{table}

We put $c^\prime_1, c^\prime_2, c^\prime_3$ into the Eq. (\ref{relative}) and plot the surface of dynamics of relative entropy of coherence for Bell-diagonal states under local nondissipative channels.

When Bell-diagonal states under four kinds of channels, as the value of relative entropy of coherence increase, surfaces of coherence become fat [see (a), (b) of Fig. 5, 6, 7, 8]. Furthermore, when $p$ increase, surfaces of coherence under BF and BPF become two opposite surface [see (c) of Fig 5, 7], surface of coherence under PF become four small triangle surface [Fig. 6(c)], and surface of coherence under GAD become cylinder [see Fig. 8(c)].

On the other hand,  for $c_{1}=-0.1, c_{2}=0.4, c_{3}=0.4$ and $c_{1}=-0.5, c_{2}=0.1, c_{3}=0.1$, the dynamic behavior of relative entropy of coherence of Bell-diagonal states under bit flip, phase flip, bit-phase flip, and generalized amplitude damping channels is depicted in Fig. 9. We find that $C_{pf}$ and $C_{gad}$ approach zero as $p$ increases and all coherence under local nondissipative channels decrease. It is a special case of the relative entropy of coherence decreasing strictly for all nontrivial evolutions\cite{Bromley}. When $C_{bf}$ decrease as $p$ increase [see Fig. 9(a)], it keep nearly unchange [see Fig. 9(b)]. Although the result like the the freezing phenomenon of coherence\cite{Bromley,Yu}, it is not all.

\begin{figure}[h]
\begin{center}
\scalebox{1.0}{(a)}{\includegraphics[width=3.5cm]{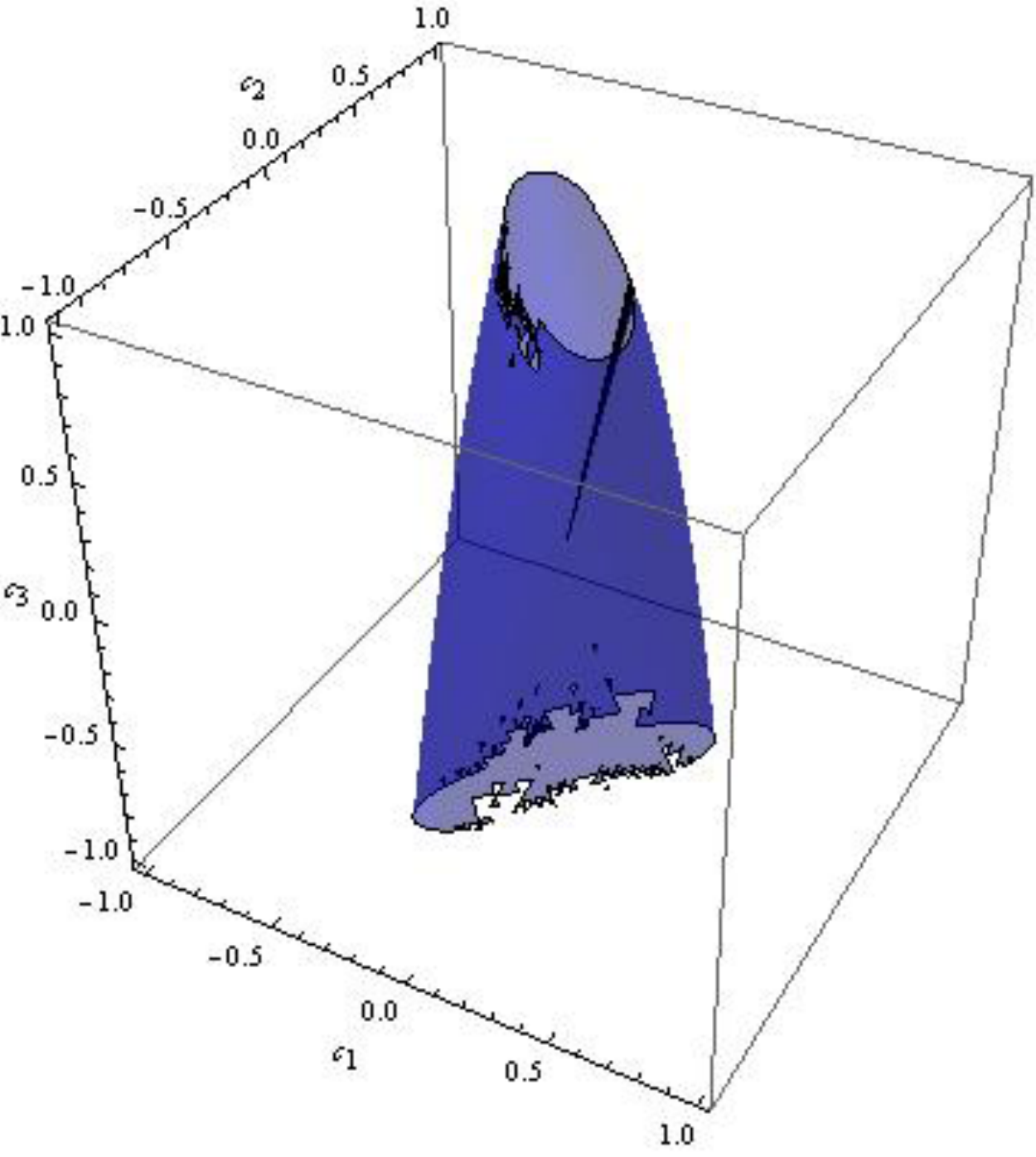}}
\scalebox{1.0}{(b)}{\includegraphics[width=3.5cm]{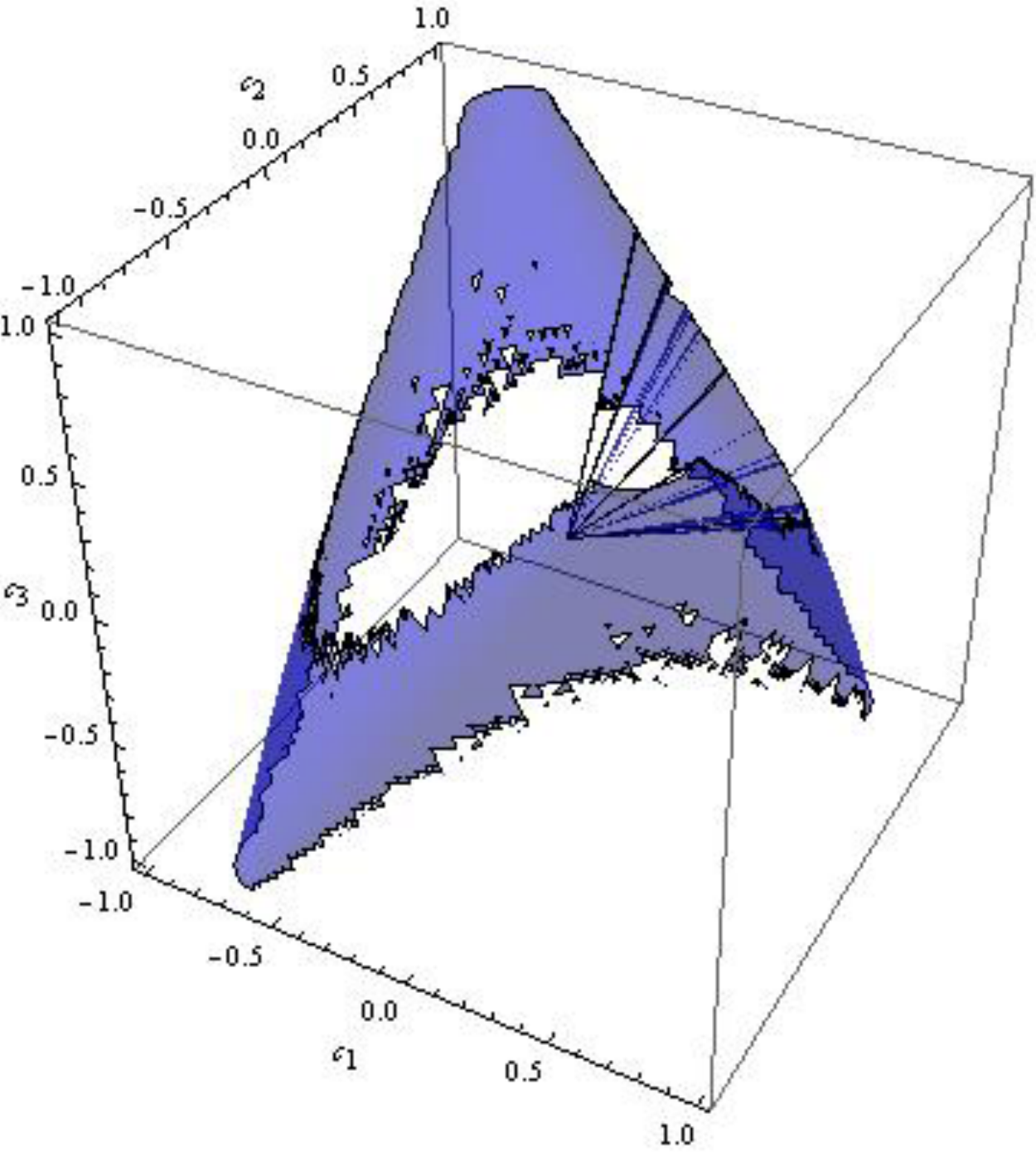}}
\scalebox{1.0}{(c)}{\includegraphics[width=3.5cm]{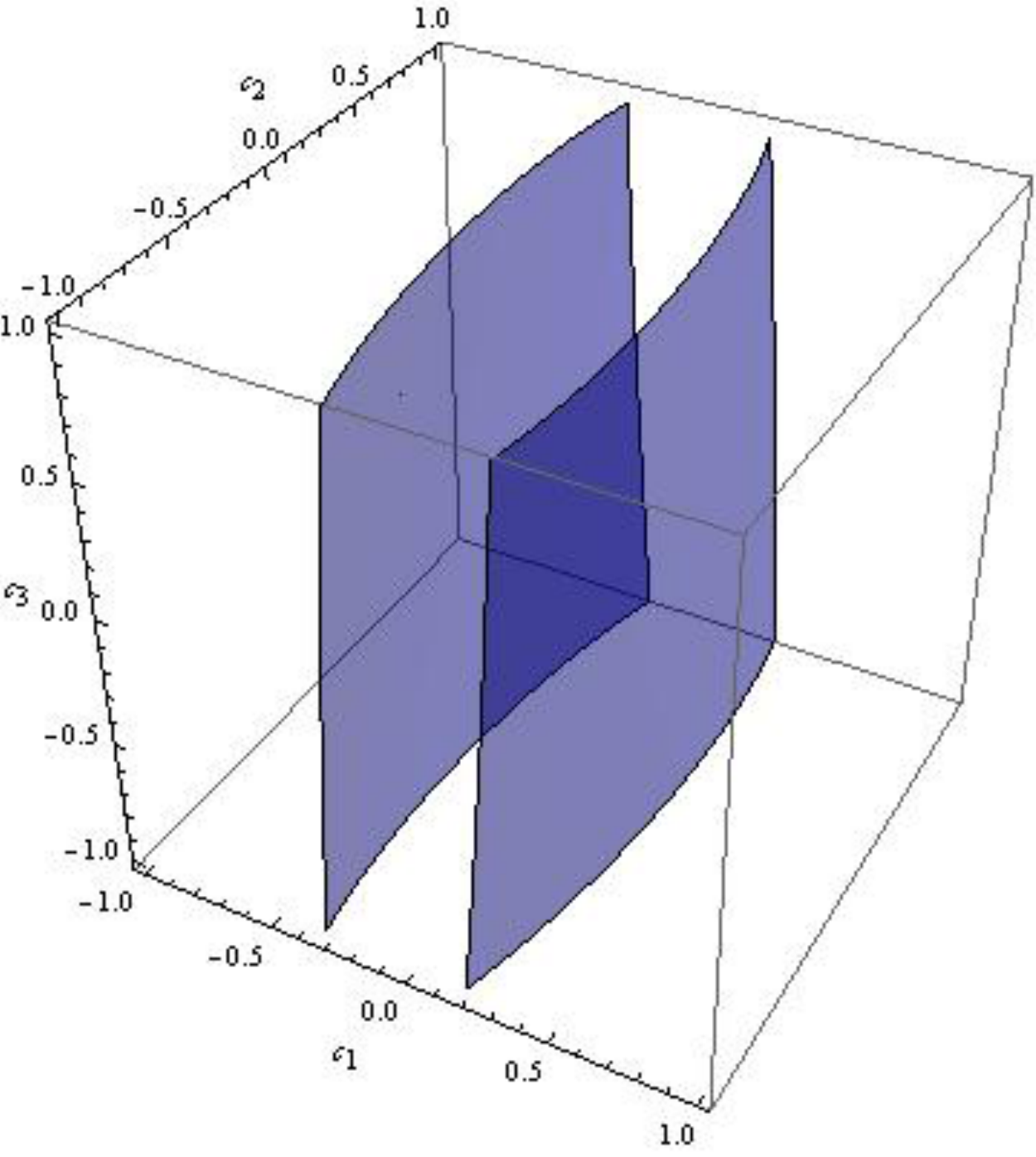}}
\caption{Surfaces of relative entropy of coherence for Bell-diagonal states under bit flip channels:(a) $p=0.1, C_{r}(\rho)=0.1$; (b) $p=0.1, C_{r}(\rho)=0.5$; (c) $p=0.5, C_{r}(\rho)=0.1$.}
\end{center}
\end{figure}

\begin{figure}[h]
\begin{center}
\scalebox{1.0}{(a)}{\includegraphics[width=3.5cm]{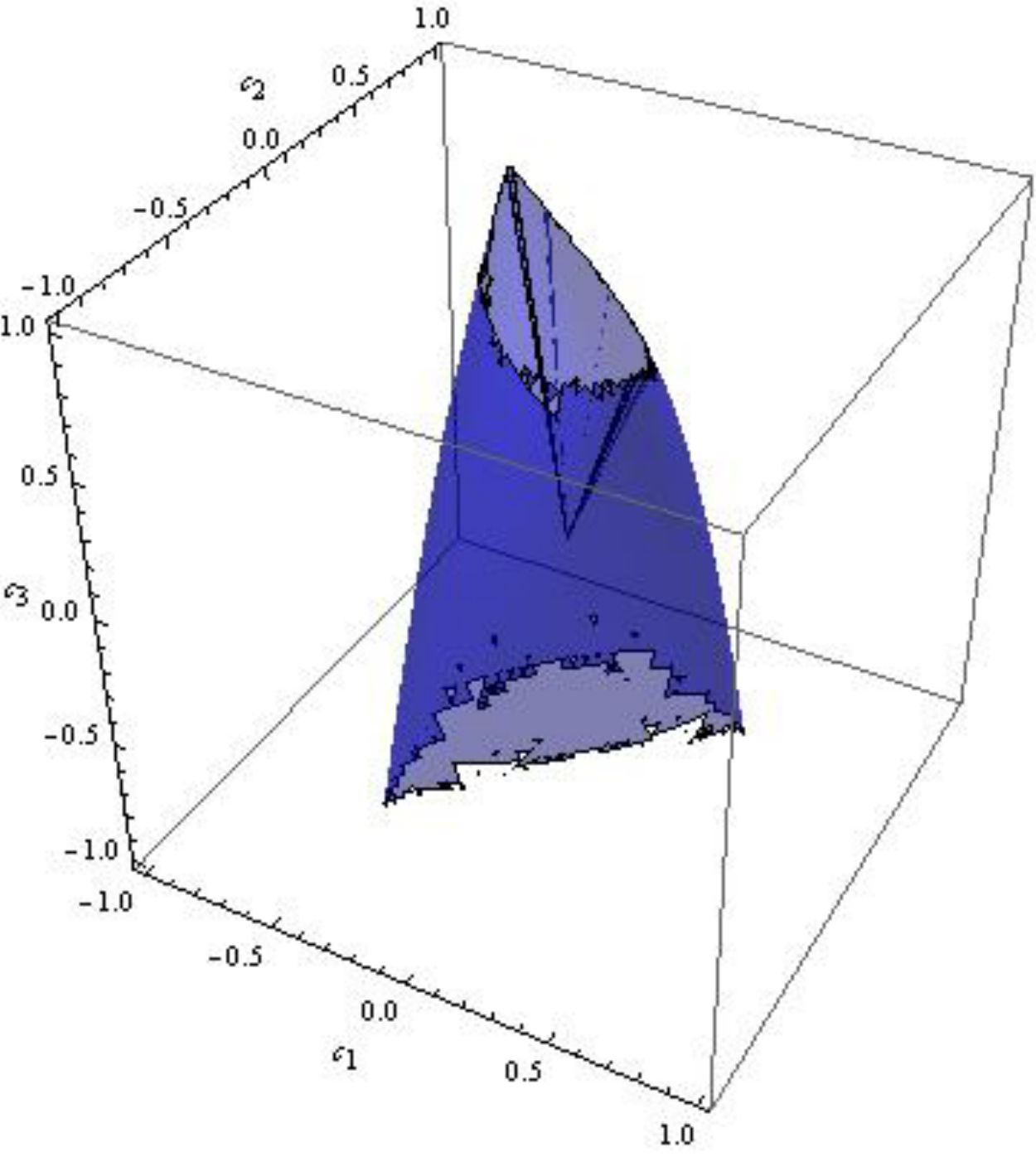}}
\scalebox{1.0}{(b)}{\includegraphics[width=3.5cm]{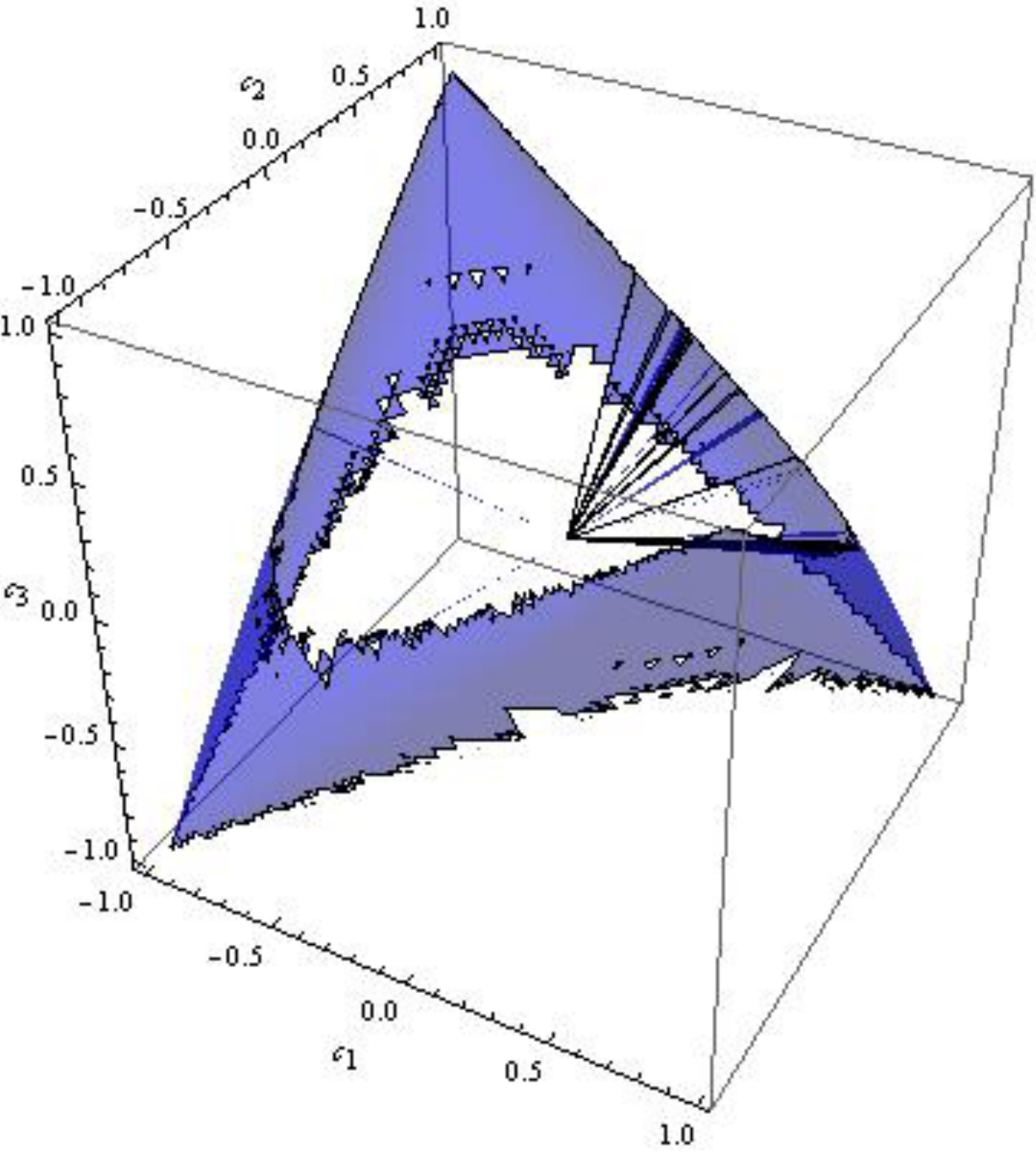}}
\scalebox{1.0}{(c)}{\includegraphics[width=3.5cm]{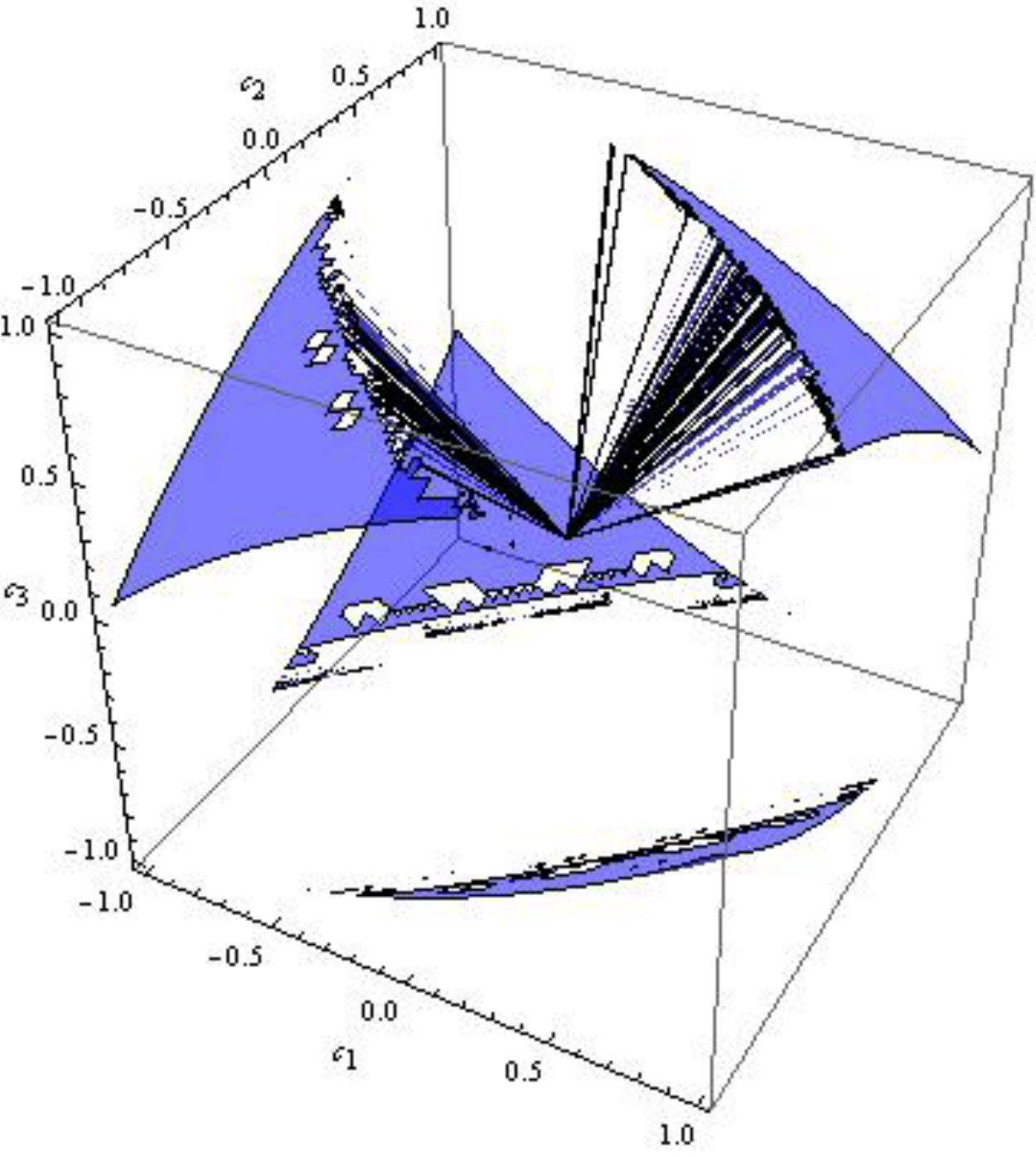}}
\caption{Surfaces of relative entropy of coherence for Bell-diagonal states under phase flip channels:(a) $p=0.1, C_{r}(\rho)=0.1$; (b) $p=0.1, C_{r}(\rho)=0.5$; (c) $p=0.5, C_{r}(\rho)=0.1$.}
\end{center}
\end{figure}

\begin{figure}[h]
\begin{center}
\scalebox{1.0}{(a)}{\includegraphics[width=3.5cm]{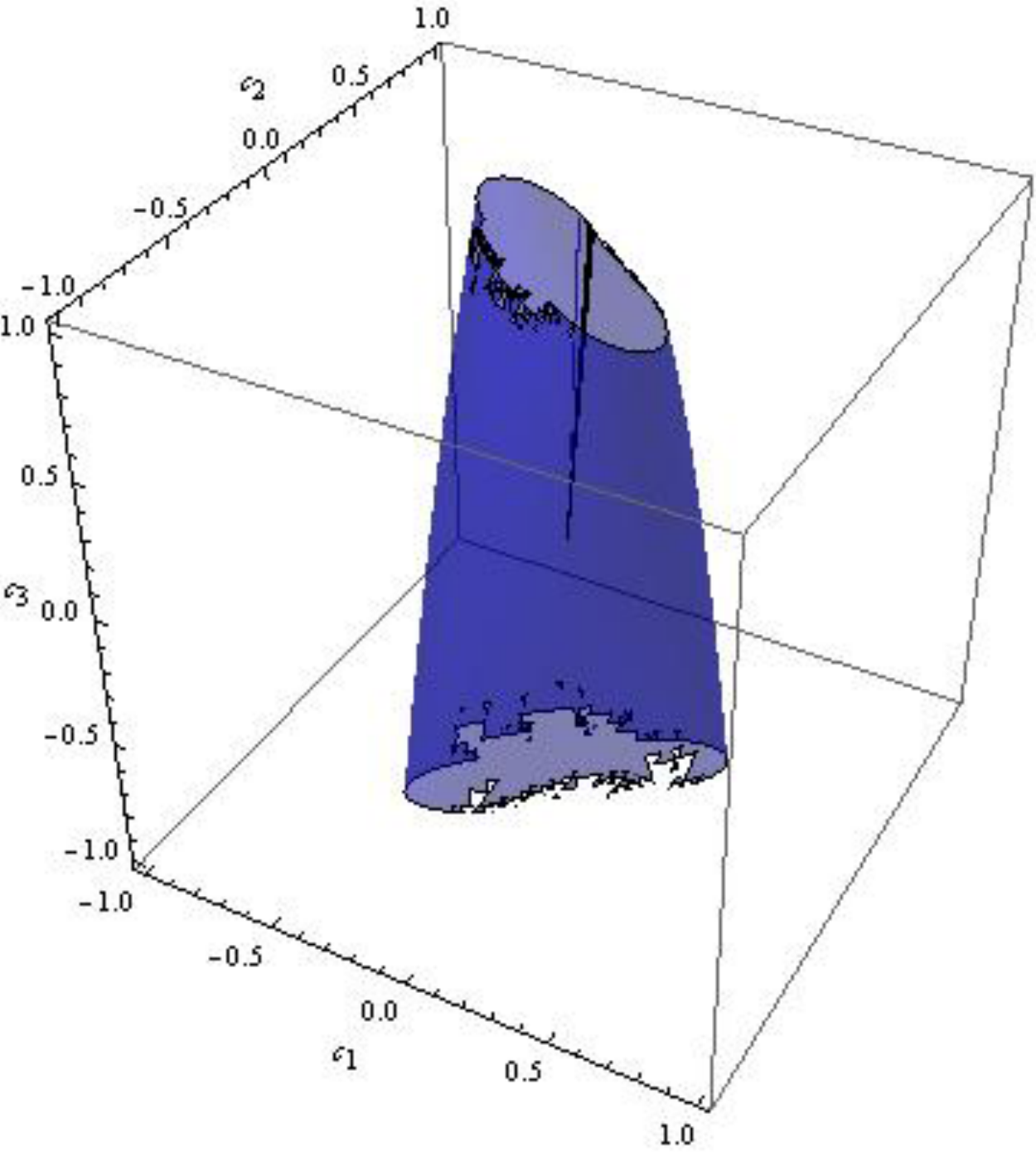}}
\scalebox{1.0}{(b)}{\includegraphics[width=3.5cm]{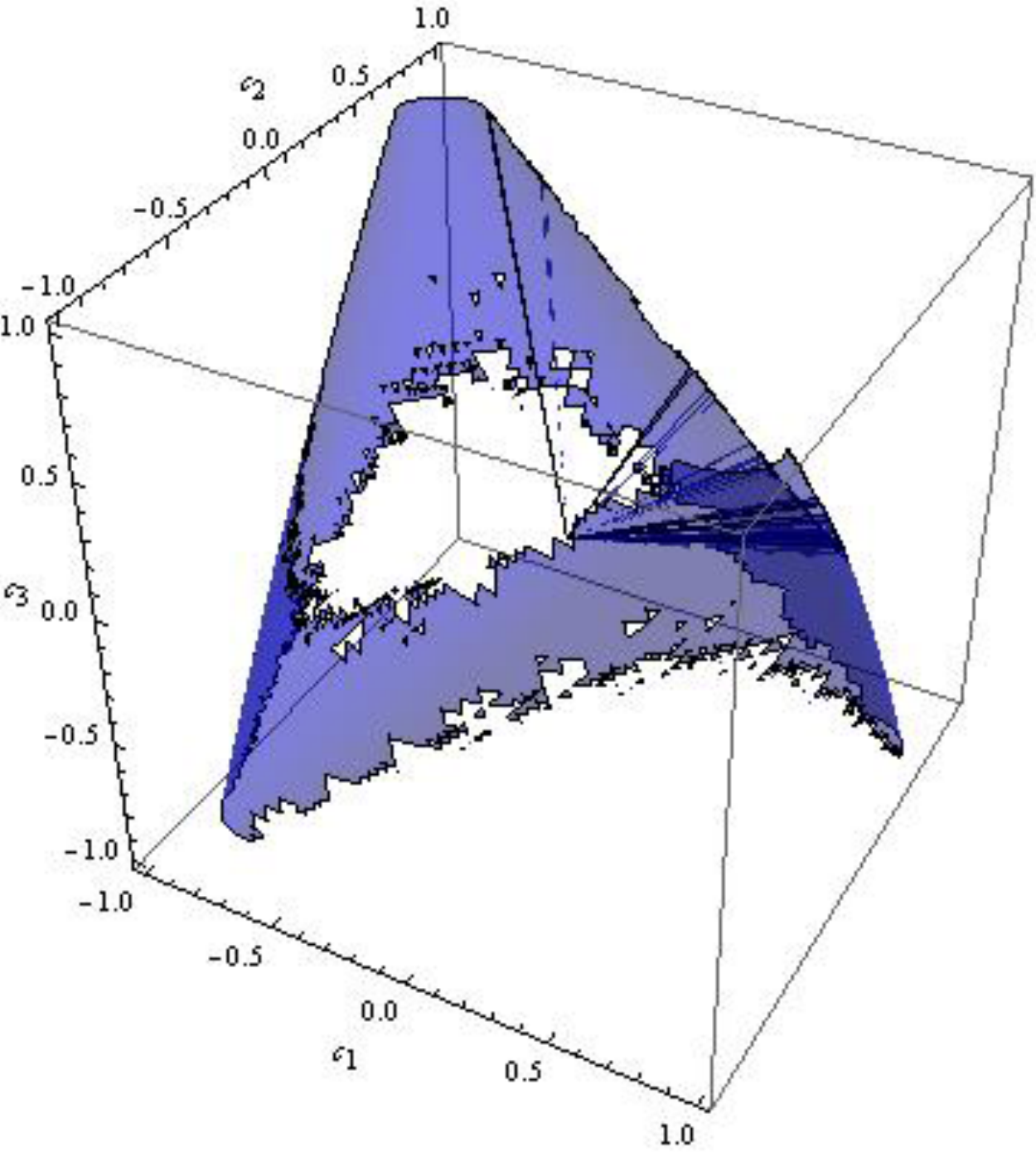}}
\scalebox{1.0}{(c)}{\includegraphics[width=3.5cm]{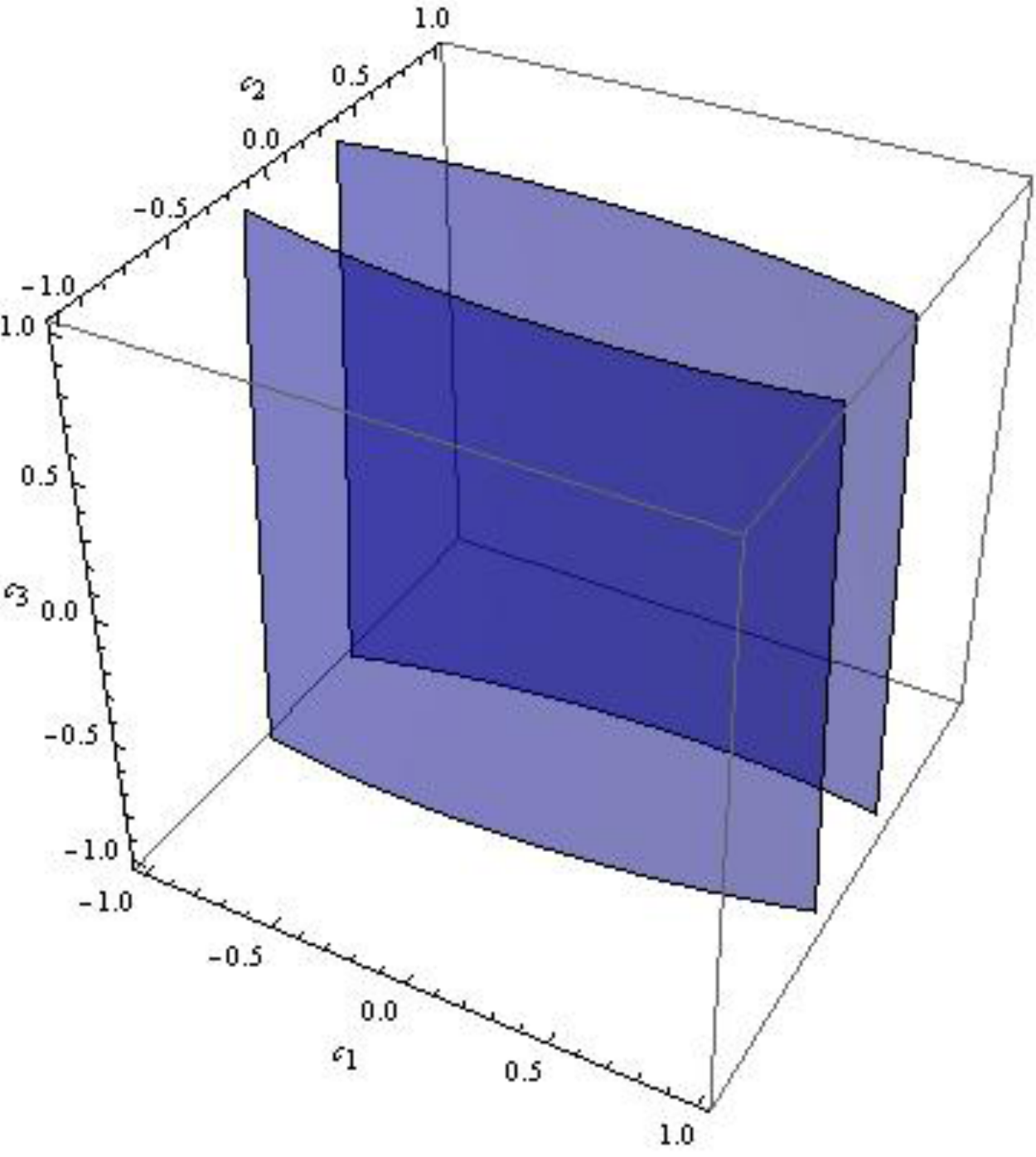}}
\caption{Surfaces of relative entropy of coherence for Bell-diagonal states under bit-phase flip channels:(a) $p=0.1, C_{r}(\rho)=0.1$; (b) $p=0.1, C_{r}(\rho)=0.5$; (c) $p=0.5, C_{r}(\rho)=0.1$.}
\end{center}
\end{figure}

\begin{figure}[h]
\begin{center}
\scalebox{1.0}{(a)}{\includegraphics[width=3.5cm]{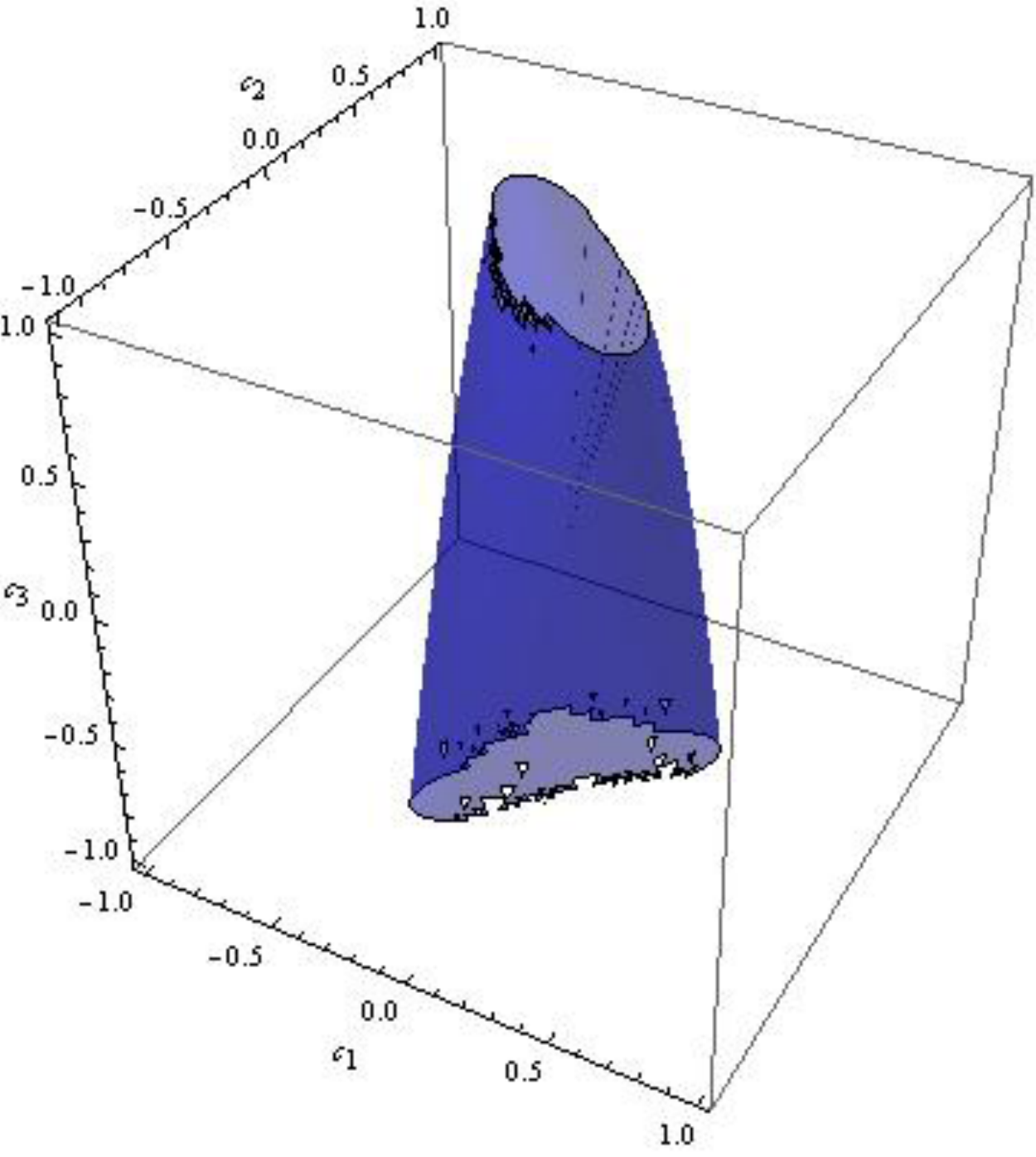}}
\scalebox{1.0}{(b)}{\includegraphics[width=3.5cm]{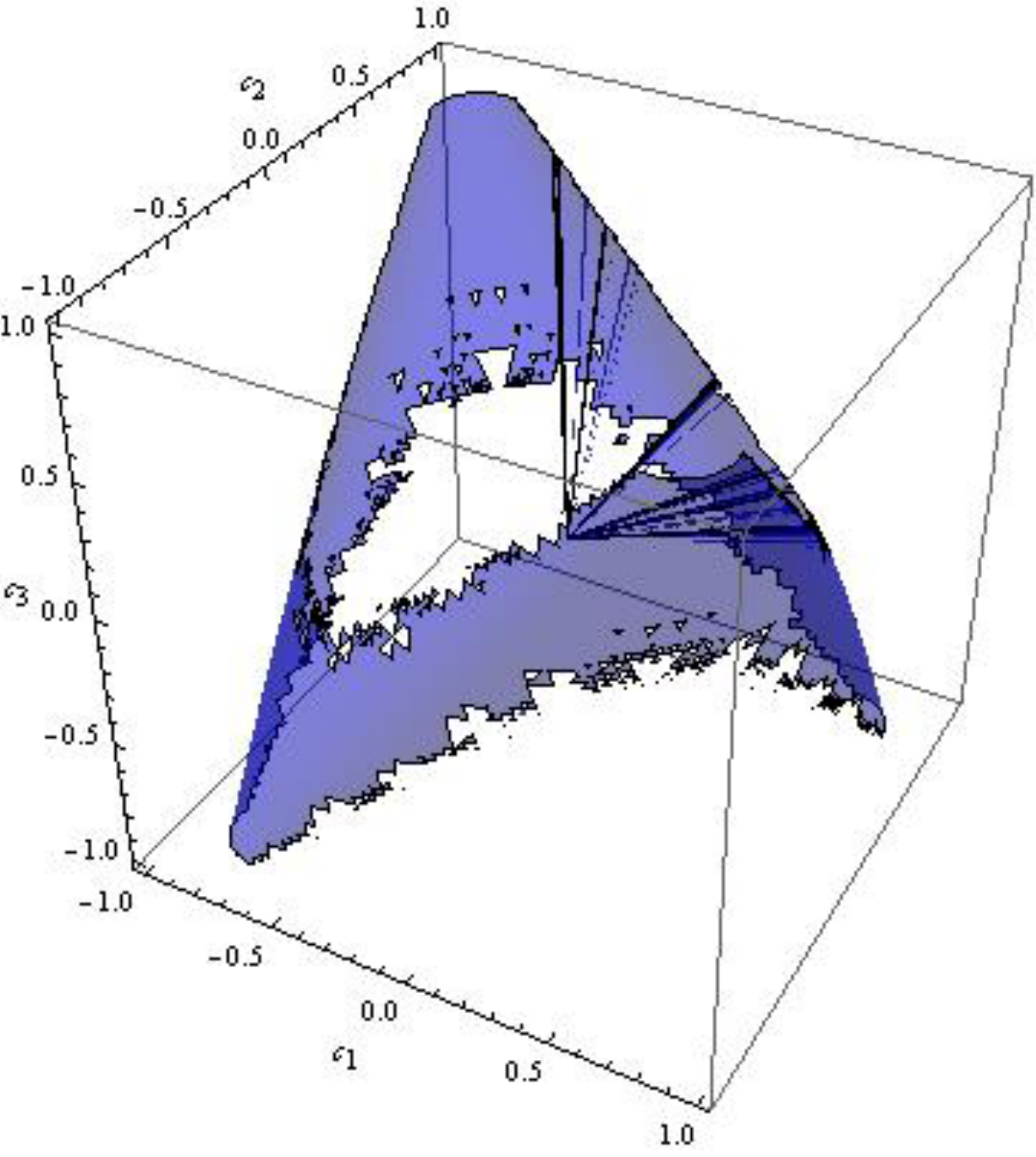}}
\scalebox{1.0}{(c)}{\includegraphics[width=3.5cm]{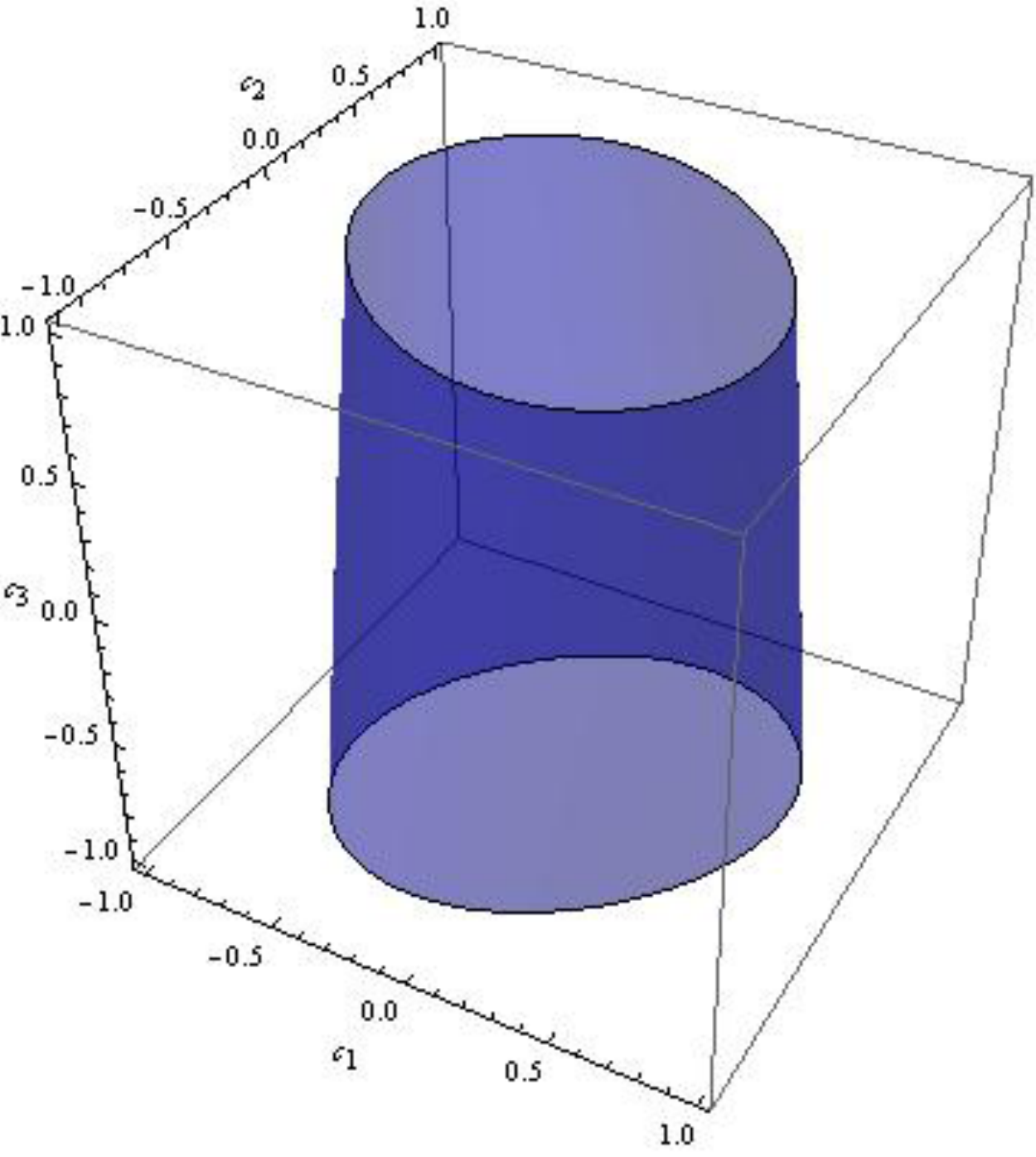}}
\caption{Surfaces of relative entropy of coherence for Bell-diagonal states under generalized amplitude damping channels:(a) $p=0.1, C_{r}(\rho)=0.1$; (b) $p=0.1, C_{r}(\rho)=0.5$; (c) $p=0.5, C_{r}(\rho)=0.1$.}
\end{center}
\end{figure}

\begin{figure}[h]
\begin{center}
\raisebox{6.2em}{(a)}\includegraphics[height=4.90cm,width=4.90cm]{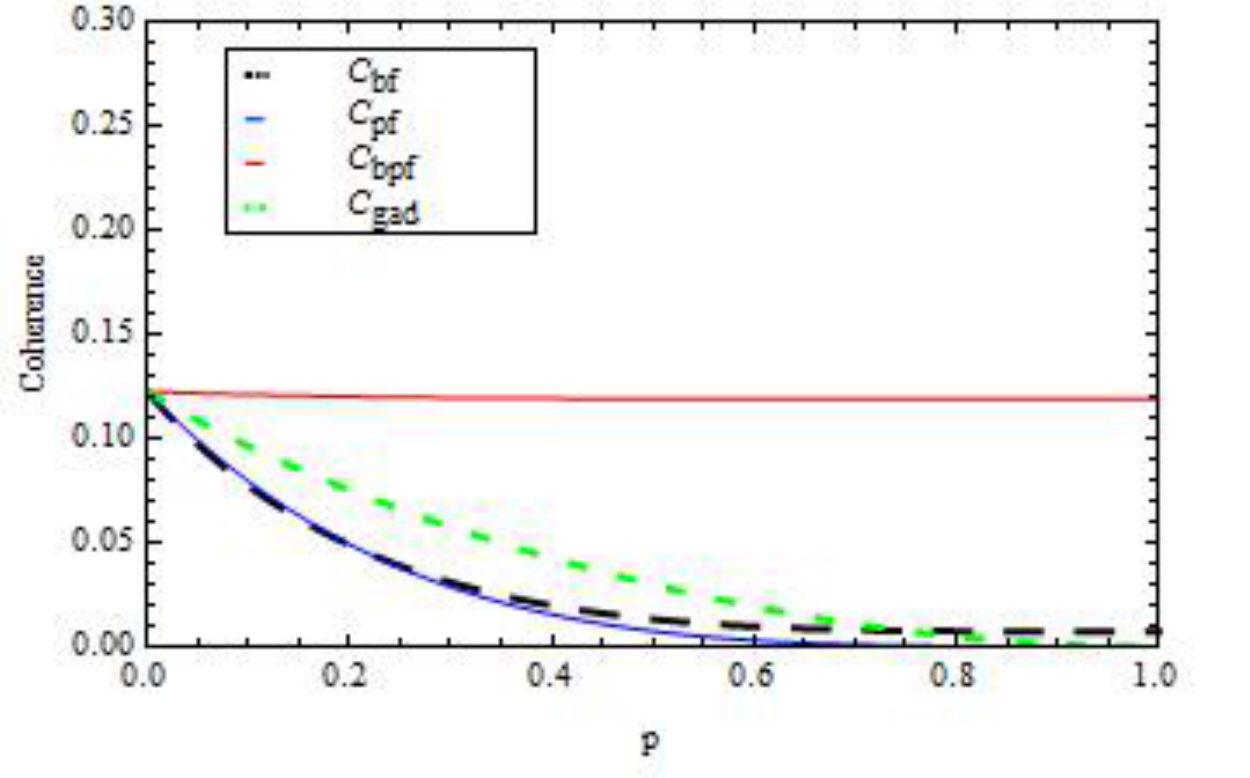}
\qquad
\raisebox{6.2em}{(b)}\includegraphics[height=4.90cm,width=4.90cm]{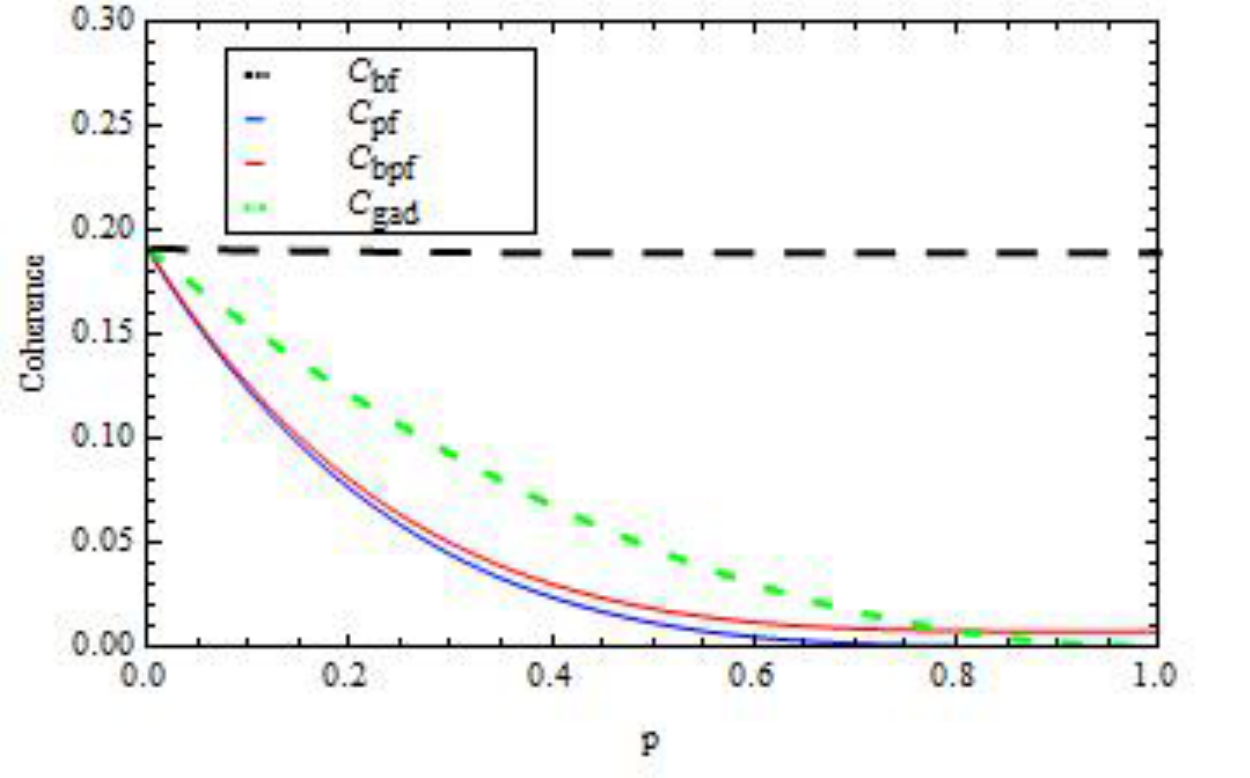}
\end{center}
\caption{(Color online) Relative entropy of coherence for Bell-diagonal states under bit flip, phase flip, bit-phase flip, and generalized amplitude damping channels  as a function of $p$ with $C_{bf}, C_{pf}, C_{bpf}, C_{gad}$  denoting respectively: (a) $c_{1}=-0.1, c_{2}=0.4, c_{3}=0.4$; (b) $c_{1}=-0.5, c_{2}=0.1, c_{3}=0.1$.}
\end{figure}

\section{\bf summary}\label{IIII}

In this work, we have calculated several distance-based quantifiers of coherence for Bell-diagonal states including $l_{1}$-norm of coherence and
relative entropy of coherence, and give the distribution of their geometry. The geometry between $l_{1}$-norm of coherence and relative entropy of coherence are basically the same except the shape of the parallelogram.  The level surface consist of the tubes running along the Cartesian axes $C_{3}$. The tubes are cut off by the state tetrahedron at their ends. As coherence decreases, the tubes collapse to the Cartesian axes $C_{3}$. As coherence increases, the tube structure is obscured. We find that when coherence equal to $0$, and the geometry of coherence is the Cartesian axes $C_{3}$ that is in the set of separable Bell-diagonal states completely. When coherence is tiny, surfaces start to move to entangled region. As coherence increase, the surfaces distribute in regions both separable and entangled. When coherence approaches 1, its surfaces almost distribute in entangled regions. When coherence equal to $1$, its surfaces is the four vertexes of the tetrahedron which is four Bell states.

We find that quantum discord equal to relative entropy of coherence for Bell-diagonal states if and only if $c_{3}=max\{|c_{1}|,|c_{2}|,|c_{3}|\}$. We plot the surfaces of the equality.

We plot surfaces of relative entropy of coherence for $X$ states. We find the surface shrinks with the effect of $r$ and $s$ and the shrinking rate becomes larger with the increasing $|r|$ and $|s|$. For larger $r$ and $s$, the picture is moved up the plane $c_{3}=0$. For larger coherence and small $r$ and $s$, the surface become fat and short.

We plot the surfaces of dynamics of relative entropy of coherence for Bell-diagonal states under local nondissipative channels. We find that when Bell-diagonal states under four kinds of channels, as the value of relative entropy of coherence increase, surfaces of coherence become fat. What is more, when $p$ increase, surfaces of coherence under bit flip and bit-phase flip channels become two opposite surface, surface of coherence under phase flip channel become four small triangle surface, and surface of coherence under generalized amplitude damping channel become cylinder.

We study the dynamic behavior of relative entropy of coherence of Bell-diagonal states. We find that all coherence under local nondissipative channels decrease and coherence under phase flip channel and generalized amplitude damping channel approach zero as $p$ increases.

\bigskip
\noindent {\bf Acknowledgments} We thank Q. Quan and T. Ma for useful discussions. This work was supported by the Science and Technology Research Plan Project of the Department of Education of Jilin Province in the Twelfth Five-Year Plan, the National Natural Science Foundation of China under grant Nos. 11175248, 11275131, 11305105.


\begin{thebibliography}{18}



\bibitem{sashki} T. Sashki, Y. Yamamoto, and M. Koashi, Nature (London) \textbf{509}, 475 (2014).
\bibitem{aberg} J. Aberg, Phys. Rev. Lett. \textbf{113}, 150402 (2014).
\bibitem{Baumgratz} T. Baumgratz, M. Cramer, and M. B. Plenio, Phys. Rev. Lett. \textbf{113}, 140401 (2014).
\bibitem{Bagan} E. Bagan, J. A. Bergou, S. S. Cottrell, and M. Hillery, Phys. Rev. Lett. \textbf{116}, 160406 (2016).
\bibitem{Jha} P. K. Jha, M. Mrejen, J. Kim, C. Wu, Y. Wang, Y. V. Rostovtsev, and X. Zhang, Phys. Rev. Lett. \textbf{116}, 165502 (2016).
\bibitem{Kammerlander} P. Kammerlander and J. Anders, Sci. Rep. \textbf{6}, 22174 (2016).
\bibitem{Giovannetti} V. Giovannetti, S. Lloyd, and L. Maccone, Science \textbf{306}, 1330 (2004).
\bibitem{Demkowicz} R. Demkowicz-Dobrza{\'n}ski and L. Maccone, Phys. Rev. Lett. \textbf{113}, 250801 (2014).
\bibitem{Giovannetti1} V. Giovannetti, S. Lloyd, and L. Maccone, Nat. Photonics \textbf{5}, 222 (2011).
\bibitem{Glauber} R. J. Glauber, Phys. Rev. \textbf{131}, 2766 (1963).
\bibitem{Sudarshan} E. C. G. Sudarshan, Phys. Rev. Lett. \textbf{10}, 277 (1963).
\bibitem{Mandel} L. Mandel and E. Wolf, \emph{Optical Coherence and Quantum Optics} (Cambridge University Press, Cambridge, UK, 1995).
\bibitem{Narasimhachar} V. Narasimhachar and G. Gour, Nat. Commun. \textbf{6}, 7689 (2015).
\bibitem{Oppenheim} P. {\'C}wikli{\'n}ski, M. Studzi{\'n}ski, M. Horodecki, and J. Oppenheim, Phys. Rev. Lett. \textbf{115}, 210403 (2015).
\bibitem{Lostaglio} M. Lostaglio, D. Jennings, and T. Rudolph, Nat. Commun. \textbf{6}, 6383 (2015).
\bibitem{Lostaglio1} M. Lostaglio, K. Korzekwa, D. Jennings, and T. Rudolph, Phys. Rev. X \textbf{5}, 021001 (2015).
\bibitem{Vazquez}H. Vazquez, R. Skouta, S. Schneebeli, M. Kamenetska, R. Breslow, L. Venkataraman, and M. S. Hybertsen, Nat. Nanotechnol. \textbf{7}, 663 (2012).
\bibitem{Wacker} O. Karlstr\"{o}m, H. Linke, G. Karlstr\"{o}m, and A. Wacker, Phys. Rev. B \textbf{84}, 113415 (2011).
\bibitem{Plenio} M. B. Plenio and S. F. Huelga, New J. Phys. \textbf{10}, 113019 (2008).
\bibitem{Rebentrost} P. Rebentrost, M. Mohseni, and A. Aspuru-Guzik, J. Phys. Chem. B \textbf{113}, 9942 (2009).
\bibitem{Lloyd} S. Lloyd, J. Phys: Conf. Ser. \textbf{302}, 012037 (2011).
\bibitem{Li} C.-M. Li, N. Lambert, Y.-N. Chen, G.-Y. Chen, and F. Nori, Sci. Rep. \textbf{2}, 885 (2012).
\bibitem{Huelga} S. F. Huelga and M. B. Plenio, Contemp. Phys. \textbf{54}, 181 (2013).
\bibitem{shao} L.-H. Shao, Z. Xi, H. Fan and Y. Li, Phys. Rev. A \textbf{91} 042120 (2015).
\bibitem{Rastegin} A. E. Rastegin, Phys. Rev. A \textbf{93} 032136 (2016).
\bibitem{Chitambar} E. Chitambar and  G. Gour, Phys. Rev. A \textbf{94} 052336 (2016).
\bibitem{Ma} J. Ma, B. Yadin, D. Girolami, V. Vedral, and M. Gu, Phys. Rev. Lett. \textbf{116}, 160407 (2016).
\bibitem{Radhakrishnan} C. Radhakrishnan, M. Parthasarathy, S. Jambulingam, and T. Byrnes, Phys. Rev. Lett. \textbf{116}, 150504 (2016).
\bibitem{Streltsov} A. Streltsov, U. Singh, H. S. Dhar, M. N. Bera, and G. Adesso, Phys. Rev. Lett. \textbf{115}, 020403 (2015).
\bibitem{Yao} Y. Yao, X. Xiao, L. Ge, and C. P. Sun, Phys. Rev. A \textbf{92}, 022112 (2015).
\bibitem{Xi} Z. Xi, Y. Li, and H. Fan, Sci. Rep. \textbf{5}, 10922 (2015).
\bibitem{Bromley} T. R. Bromley, M. Cianciaruso, and G. Adesso (2015), Phys. Rev. Lett. \textbf{114}, 210401 (2015).
\bibitem{Yu} X.-D. Yu, D.-J. Zhang, C. L. Liu, and D. M. Tong, Phys. Rev. A \textbf{93}, 060303 (2016).
\bibitem{Horodecki} R. Horodecki, P. Horodecki, M. Horodecki, and K. Horodecki, Rev. Mod. Phys. \textbf{81}, 865 (2009), and references therein.
\bibitem{Horodecki1} R. Horodecki and M. Horodecki, Phys. Rev. A \textbf{54}, 1838 (1996).
\bibitem{Lang} M. D. Lang and C. M. Caves, Phys. Rev. Lett. \textbf{105}, 150501 (2010).
\bibitem{Li1} B. Li, Z.-X. Wang, and S.-M. Fei, Phys. Rev. A \textbf{83}, 022321 (2011).
\bibitem{wang} Y.-K. Wang, T. Ma, B. Li, Z.-X. Wang, Commun. Theor. Phys. \textbf{59}, 540 (2013).
\bibitem{wang1} Y.-K. Wang, T. Ma, H. Fan, S.-M. Fei, Z.-X. Wang, Quantum Inf Process  \textbf{13}, 283 (2014).
\bibitem{quan} Q. Quan, H. Zhu, S.-Y. Liu, S.-M. Fei, H. Fan, and W.-L. Yang, Sci. Rep. \textbf{6}, 22025 (2016).

\bibitem{Rana} S. Rana, P. Parashar, and M. Lewenstein, Phys. Rev. A \textbf{93}, 012110 (2016).
\bibitem{luo} S. Luo,  Phys. Rev. A \textbf{77}, 042303 (2008).
\bibitem{chen} Q. Chen, C. Zhang, S. Yu, X. X. Yi, and C. H. Oh, Phys. Rev. A \textbf{84}, 042313 (2011).
\bibitem{nielsen} M. A. Nielsen, and I. L. Chuang,  \emph{Quantum Computation and Quantum Information} (Cambridge University Press,  Cambridge,  UK,  2000).
\bibitem{mont} J. D. Montealegre, F. M. Paula, A. Saguia, and M. S Sarandy, Phys. Rev. A \textbf{87}, 042115 (2013).




\end{thebibliography}
\end{document}